\documentclass[aps,
superscriptaddress,
twocolumn,
amsmath,amssymb,
prc,
floatfix,
preprintnumbers,
nofootinbib,
]{revtex4-1}

\usepackage{graphicx}
\usepackage{dcolumn}
\usepackage{bm}
\usepackage{color}
\usepackage{hyperref}
\usepackage{graphicx}
\usepackage{braket}
\usepackage{amsmath,amssymb}
\usepackage[normalem]{ulem}
\usepackage{xcolor}

\begin{document}

\preprint{LA-UR-24-27320}

\title{Emulators for Scarce and Noisy Data:\\
Application to Auxiliary-Field Diffusion Monte Carlo for Neutron Matter
}

\author{Cassandra~L.~Armstrong}
\affiliation{Intelligence and Space Research Division, Los Alamos National Laboratory, Los Alamos, New Mexico 87545, USA}

\author{Pablo~Giuliani}
\affiliation{Facility for Rare Isotope Beams, Michigan State University, East Lansing, Michigan 48824, USA}

\author{Kyle~Godbey}
\affiliation{Facility for Rare Isotope Beams, Michigan State University, East Lansing, Michigan 48824, USA}

\author{Rahul~Somasundaram}
\affiliation{Theoretical Division, Los Alamos National Laboratory, Los Alamos, New Mexico 87545, USA}
\affiliation{Department of Physics, Syracuse University, Syracuse, New York 13244, USA}

\author{Ingo Tews}
\email{itews@lanl.gov}
\affiliation{Theoretical Division, Los Alamos National Laboratory, Los Alamos, New Mexico 87545, USA}

\date{\today}

\begin{abstract}
Understanding the equation of state (EOS) of pure neutron matter is necessary for interpreting multimessenger observations of neutron stars.
Reliable data analyses of these observations require well-quantified uncertainties for the EOS input, ideally propagating uncertainties from nuclear interactions directly to the EOS.
This, however, requires calculations of the EOS for a prohibitively large number of nuclear Hamiltonians, solving the nuclear many-body problem for each one.
Quantum Monte Carlo methods, such as auxiliary-field diffusion Monte Carlo (AFDMC), provide precise and accurate results for the neutron matter EOS, but they are very computationally expensive, making them unsuitable for the fast evaluations necessary for uncertainty propagation.
Here, we employ parametric matrix models to develop fast emulators for AFDMC calculations of neutron matter and use them to directly propagate uncertainties of coupling constants in the Hamiltonian to the EOS.
As these uncertainties include estimates of the effective field theory truncation uncertainty, this approach provides robust uncertainty estimates for use in astrophysical data analyses.
This Letter will enable novel applications such as using astrophysical observations to put constraints on coupling constants for nuclear interactions.
\end{abstract}

\maketitle


{\it Introduction.}
The equation of state (EOS) of pure neutron matter (PNM) is a key input to constrain properties of neutron stars (NSs), such as their masses, radii, and tidal deformabilities~\cite{Lattimer:2000nx,Ozel:2016oaf,Lattimer:2021emm,Chatziioannou:2024tjq}. 
This is because neutron-star properties are dominated by the neutron-star core where proton fractions are small.
Using observatories such as the Laser Interferometer Gravitational-Wave Observatory (LIGO)~\cite{LIGOScientific:2017vwq,De:2018uhw} or NASA's Neutron Star Interior Composition Explorer (NICER)~\cite{Riley:2019yda,Riley:2021pdl,Miller:2019cac,Miller:2021qha,Choudhury:2024xbk}, one can extract neutron-star properties, and hence, constrain the EOS~\cite{Capano:2019eae,Dietrich:2020efo,Landry:2020vaw,Tan:2020ics,Miller:2021qha,Raaijmakers:2021uju,Legred:2021hdx,Gorda:2022jvk,Koehn:2024set,Rutherford:2024srk}. 
The coming decades will provide even more astrophysical data from current and next-generation observatories, such as Cosmic Explorer~\cite{Reitze:2019iox,Evans:2021gyd} and the Einstein Telescope~\cite{Punturo:2010zz,Maggiore:2019uih}, ushering in a data-rich era for neutron stars that will allow us to further constrain the EOS. 
To properly analyze and extract robust nuclear-physics information from this multimessenger data, it is important to provide statistically meaningful EOS priors with a strong foundation in uncertainty quantification.

At low densities, the neutron-matter EOS can be calculated using nuclear Hamiltonians from chiral effective field theory (EFT).
Chiral EFT is a systematic expansion for nuclear forces and describes nuclear interactions in neutron matter up to densities of about two times the nuclear saturation density, $n_{\rm sat} \sim 0.16$~fm$^{-3}$~\cite{Tews:2018kmu,Essick:2020flb,Drischler:2020hwi,saturation2024}. 
Chiral EFT Hamiltonians depend on a set of parameters, called low-energy couplings (LECs), that are usually calibrated to scattering data and properties of atomic nuclei~\cite{Navratil:2007we,Ekstrom:2015rta,Reinert:2017usi,Entem:2017gor,somasundaram2025emulators}.
Using chiral EFT Hamiltonians, one can then employ a quantum many-body method to solve for the EOS of PNM.
Here, we employ accurate and precise quantum Monte Carlo (QMC) methods~\cite{Carlson:2014vla}, and in particular, auxiliary-field diffusion Monte Carlo (AFDMC)~\cite{Schmidt:1999lik}, to obtain the energy per particle in PNM. 
Importantly, the use of chiral EFT interactions enables uncertainty estimates due to the truncation of the EFT expansion, reflecting our incomplete knowledge of nuclear interactions.~\cite{Epelbaum:2014efa,Drischler:2020hwi}.
Currently, these uncertainties are estimated \textit{a posteriori}: calculations are performed order-by-order in the EFT expansion, and an uncertainty band is estimated from these results~\cite{Epelbaum:2014efa,Drischler:2020hwi}.
Recent AFDMC results for the EOS of PNM have employed such {\it a posteriori} estimates for theoretical uncertainties~\cite{Tews:2024owl}.
Nevertheless, the statistical interpretation of these error bands is either unclear~\cite{Epelbaum:2014efa} or distributions are assumed to be Gaussian~\cite{Drischler:2020hwi}.

\begin{figure}[t]
    \centering
    \includegraphics[width=1.0\columnwidth]{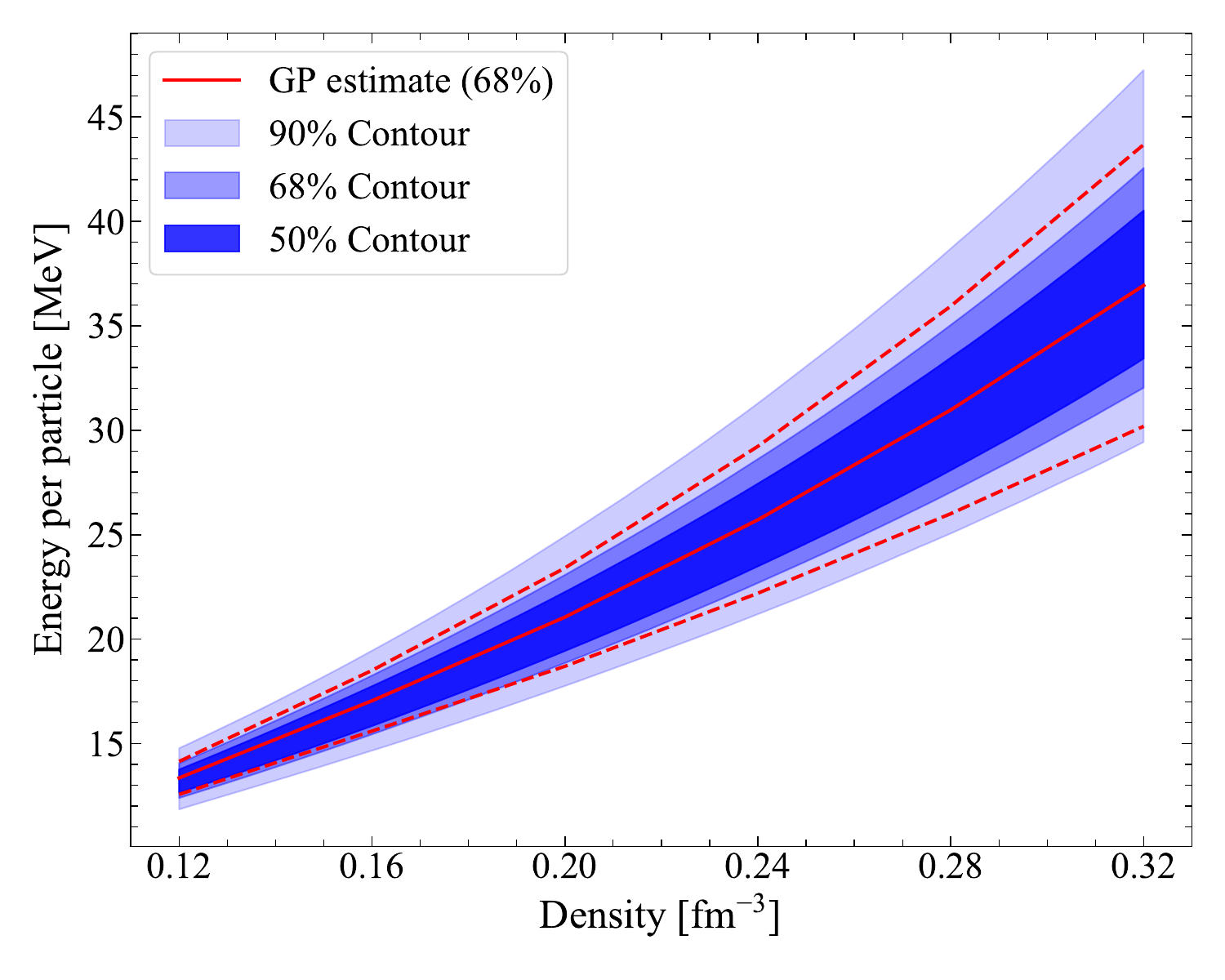}
    \caption{Full propagated error bands for the energy per particle of PNM as function of number density at 68\% and 90\% confidence intervals.
    The red band shows the mean and 68\% uncertainty of Ref.~\cite{Tews:2024owl} using a Gaussian Process error prescription~\cite{Drischler:2020hwi}.
    }
    \label{fig:EOS_distribution}
\end{figure}

The truncation uncertainty of the EFT expansion can also be accounted for during the fit of the LECs~\cite{Somasundaram:2023sup}.
Then, the distributions of the LECs include not only statistical uncertainties stemming from the fit data, but also an estimate for the systematic EFT uncertainty.
A robust statistical interpretation of such uncertainty bands can also be obtained by propagating these uncertainties from the LECs~\cite{Somasundaram:2023sup} directly to the EOS~\cite{Jiang:2022tzf,Jiang:2022oba}.
In the case of the EOS of PNM, this would allow us to extract a consistent shape of the EOS uncertainty distribution at each density given an ansatz for the distribution of the LECs.
This approach, however, is currently not feasible for the EOS of PNM because AFDMC calculations are very computationally expensive.
For example, the calculation of the energy per particle at one fixed density and for one LEC set costs on the order of several 100,000 CPU-hours.
In this Letter, we address this problem and present emulators for AFDMC calculations of pure neutron matter based on the parametric matrix model (PMM)~\cite{Cook:2024toj,somasundaram2025emulators}.
These emulators allow us to speed up  AFDMC calculations by a factor of $10^8$, enabling us to rapidly carry out the many thousands of calculations necessary for error propagation from LECs to the EOS.
Using these emulators, we provide reliable uncertainty bands for the EOS of PNM, and show our main result in Fig.~\ref{fig:EOS_distribution}.

{\it Methods---}The PMM~\cite{Cook:2024toj} is a general purpose machine-learning algorithm that has shown excellent performance on various tasks, including image classification, curve extrapolation, and surrogate modeling~\cite{somasundaram2025emulators,reed2024toward}.
For the latter, it has been especially powerful for emulating eigenvalue equations such as the Schr{\"o}dinger equation. 
In this context, it can be interpreted as a data-driven variant of the reduced-basis method that has been recently adopted by the low-energy nuclear-theory community~\cite{Duguet:2023wuh,giuliani2023bayes,drischler2023buqeye,bonilla2022training, cheng2024reduced, odell2024rose,frame2018eigenvector,hu2022ab}. 
The reduced-basis method identifies the physically relevant subspace spanned by solutions for different LECs through a basis expansion informed by previous high-fidelity solutions---solutions found in the full model space that are here given by highly precise and accurate results from AFDMC---and then projects the necessary equations describing the system into this subspace.
In contrast, the PMM approach implicitly identifies a relevant effective low-dimensional space by exploiting the eigensystem of matrices whose elements are fit to data, which here are the high-fidelity AFDMC energies.
In this Letter, we construct a PMM inspired by the structure of the chiral EFT Hamiltonian up to a chosen order in the EFT expansion, that depends on the LECs in an affine way.
Generally, we define 
\begin{equation}
\hat{H}(\{c_i\}) = H_0 + c_1 \cdot H_1+c_2 \cdot H_2+\cdots\,,
\label{eq:PMM}
\end{equation}
where
$\{c_i\}$ are our control parameters, i.e., the LECs of the Hamiltonian, and all the involved matrices have dimensions $\text{N}_\text{dim}\times \text{N}_\text{dim}$. For example, in the $\text{N}_\text{dim}=2$ case and for two control parameters $c_1$ and $c_2$ we have:
\begin{equation}
    \hat{H}=\begin{bmatrix}
                    \alpha & \beta \\
                    \beta & \gamma
                \end{bmatrix}, 
                H_0=\begin{bmatrix}
                    a & 0 \\
                    0 & b
                \end{bmatrix}, 
                H_1=\begin{bmatrix}
                    c & d \\
                    d & e
                \end{bmatrix},
                H_2=\begin{bmatrix}
                    f & g \\
                    g & h
                \end{bmatrix}\,,
                \label{eq:PMM_matrix}
\end{equation}
with real matrix elements $a-h$,
with $H_0$ being a diagonal matrix containing all terms that are not dependent on the LECs, and $H_1$ and $H_2$ being symmetric matrices. 
The matrix elements are found through fitting a selected eigenvalue of $\hat{H}$ to data for the observable we want to emulate, in this case, AFDMC calculations of the PNM energy.
The matrices in Eq.~\eqref{eq:PMM_matrix} can be extended to higher matrix dimensions N$_{\rm dim}$, which can improve the PMM emulator at the cost of more PMM parameters to fit.
This general structure for the PMM was chosen based on previous experience with the deuteron~\cite{somasundaram2025emulators} and with a toy problem discussed in the Supplemental Material~\cite{supp}.

To generate the training data we employ the AFDMC method.
AFDMC is a QMC algorithm that solves the Schr{\"o}dinger equation by performing an evolution of a trial wave function $\ket{\psi_T}$ in imaginary time $\tau$ to project out the ground state of the system $\ket{\psi_0}$~\cite{Carlson:2014vla,Lynn:2019rdt},
\begin{equation}
    \ket{\psi_0} = \lim_{\tau \to \infty} e^{-H \tau} \ket{\psi_T}\,.
\end{equation}
The direct computation of the propagator $\exp{\{-H\tau\}}$ for arbitrary $\tau$ is typically impossible, but the propagator can be calculated for small imaginary times $\delta\tau=\tau/N$ with large $N$ using Monte Carlo methods~\cite{Carlson:2014vla}.
For this, the nuclear input Hamiltonian has to be local, as is the case for the various chiral EFT interactions that have been developed in the past decades for use in QMC methods~\cite{Gezerlis:2013ipa,Gezerlis:2014zia,Piarulli:2014bda,Tews:2015ufa,Lynn:2015jua,Piarulli:2016vel,Piarulli:2017dwd,Piarulli:2019cqu,Somasundaram:2023sup}.
Using perturbation theory, however, it is also possible to treat small nonlocalities~\cite{Curry:2024gcz}.
Here, we use the local high-cutoff chiral EFT interactions of Ref.~\cite{Somasundaram:2023sup} with $R_0=0.6$~fm at leading order (LO) with 2 LECs, next-to-leading order (NLO), and next-to-next-to-leading order (N$^2$LO), with 9 LECs each.
These interactions were fit to nucleon-nucleon scattering data over a range of cutoffs using a Bayesian fitting protocol that explicitly accounts for truncation uncertainties of the EFT expansion and provides us with full posterior distributions for all LECs.
The three-nucleon interactions at N$^2$LO are determined by pion-nucleon LECs that we take from the Roy-Steiner analysis~\cite{Hoferichter:2015hva} and do not vary here.
In pure neutron matter, the shorter-range three-nucleon interactions do not contribute for nonlocal interactions~\cite{Hebeler:2009iv}, and are negligible for local interactions at high cutoffs~\cite{Tews:2024owl}.
For these interactions, we generated full AFDMC solutions for pure neutron matter for several LEC sets.
At LO (NLO), we performed AFDMC simulations with 14 neutrons at nuclear saturation density for 30 (60) LEC sets.
At N$^2$LO, we have calculated 30 high-fidelity data points using simulations with 66 neutrons at $0.12$~fm$^{-3}$, $0.16$~fm$^{-3}$, $0.24$~fm$^{-3}$, $0.32$~fm$^{-3}$, and $0.48$~fm$^{-3}$, following the final setup of Ref.~\cite{Tews:2024owl}.
Importantly, the chosen cutoff $R_0=0.6$~fm removes regulator artifacts~\cite{Lynn:2015jua,Huth:2017wzw}.
The generation of all training data cost $\sim 22$~million CPU-hours.

{\it PMMs for AFDMC---}For propagation of LEC distributions to PNM, full AFDMC simulations are prohibitively expensive. 
Here, we apply the PMM, focusing on the lowest eigenvalue, to construct a surrogate model of AFDMC instead.
We constructed PMMs at LO and NLO to understand the best setup and study the necessary amount of training data to achieve a good reproduction of AFDMC energies.
We present more details for the PMM at these orders in the Supplemental material~\cite{supp} but summarize the most important results here.
At LO, where the PMM is described by the three terms in Eq.~\eqref{eq:PMM}, we split the 30 high-fidelity data points into a training set with 20 points and validation set with the remaining 10 points.
The matrix elements of the PMM are then fit to the training energies using \textsc{SciPy}'s Dual Annealing optimization routine~\cite{2020SciPy-NMeth}, and the resulting PMM is used to predict the PNM energies for the validation LEC sets.
We note that, especially for larger N$_{\rm dim}$, we typically encounter a situation with a larger number of unknowns than training data.
The Dual Annealing approach allows us to find solutions to this ill-posed problem that nevertheless have good extrapolation power.
However, when initially setting up the PMM, we found that, instead of the error decreasing or leveling off with an increasing number of training points N$_{\rm train}$, there appeared spikes with unexpectedly high percent errors for some values of N$_{train}$.
This increase in percent error is caused by a level crossing in the resulting PMM fit that occurs at energies above those in the training set.
In these cases, the training data is described by the lowest eigenvalue of $\mathcal{\hat{H}}$, but the validation data is described in part by the second eigenvalue.
To circumvent this problem and avoid the level crossing in the PMM fit, we tested various fitting strategies.
The fitting approach that led to the best and most consistent results is defined by ensuring that the maximum energy in the training set is always included when training, regardless of the number of training points, and that the fit for N$_{\rm train}+1$ training points starts from the result of the fit for N$_{\rm train}$. 
We provide more details in the Supplemental Material~\cite{supp}.
With the specified fitting routine, we tested the validation error for the PMM as a number of different N$_{\rm train}$ and N$_{\rm dim}$. 
At LO, for N$_{\rm dim}=2$ the PMM improves as a function of N$_{\rm train}$ until N$_{\rm train} = 4$ where the error settles around $0.03$\%.
The error is similarly small for different N$_{\rm dim}>2$.

\begin{figure}[t]
    \centering    
    \includegraphics[width=1.0\columnwidth]{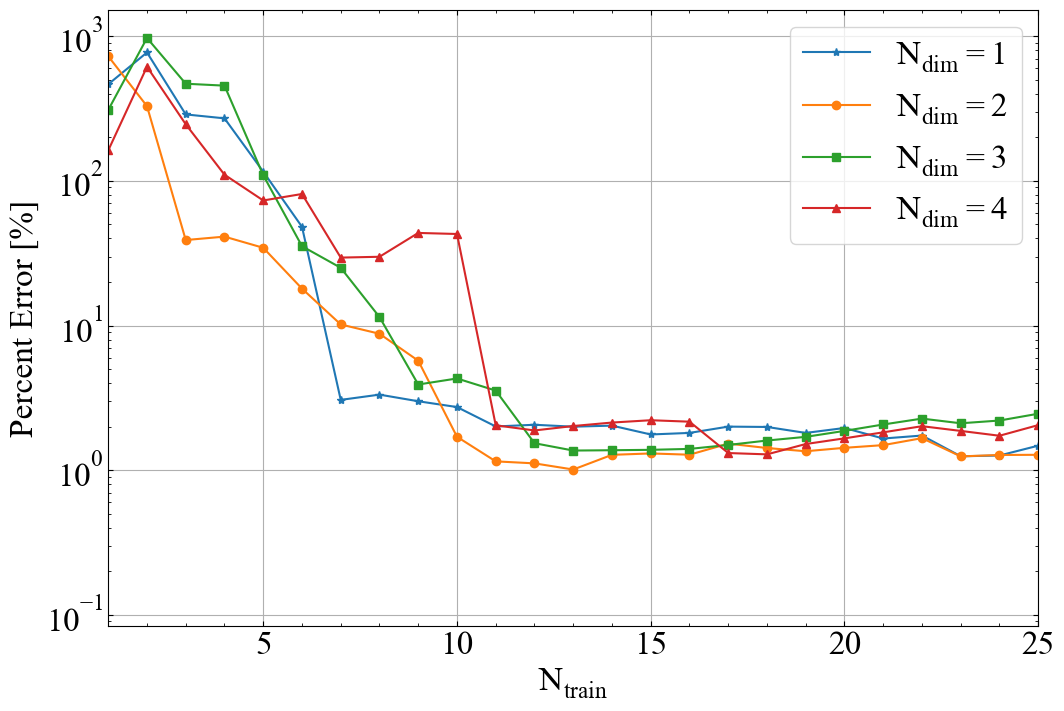}
    \caption{Convergence of the relative error of the PMM reproduction of the AFDMC validation energies with the number of training points N$_{\rm train}$ for N$_{\rm dim}=1-4$ at N$^2$LO at nuclear saturation density, $n_{\rm sat}=0.16 \; \rm{fm^{-3}}$. 
    The training and validation sets are the same for all N$_{\rm dim}$.
    }
    \label{fig:error_training_lsq_n2lo}
\end{figure}

At NLO and N$^2$LO, the number of LECs describing the nuclear Hamiltonian increases from 2 to 9.
Using the fitting protocol specified above, we first built PMMs at NLO as calculations are cheaper.
At NLO, we compared the performance of a PMM with 9 operator LECs and a PMM with the 6 relevant spectral LECs, see Ref.~\cite{Gezerlis:2014zia} for details.
With 9 operator LECs, the PMM at NLO  is set up as
\begin{equation}
    \begin{split}
        \hat{H} = H_0 &+ C_S\cdot H_1 + C_T \cdot H_2 + C_1\cdot H_3 \\ 
        &+ C_2\cdot H_4 + C_3\cdot H_5 + C_4 \cdot H_6 \\ 
        &+ C_5\cdot H_7 + C_6 \cdot H_8 + C_7\cdot H_9\,,
    \end{split}
    \label{eq:operator}
\end{equation}
and with 6 spectral LECs, the PMM is 
\begin{equation}
    \begin{split}
         \hat{H}= H_0 &+ d_{11}\cdot H_1 + d_{22} \cdot H_2  + d_3 \cdot H_3 \\ 
         &+ d_4\cdot H_4 + d_6 \cdot H_5 + d_7 \cdot H_6\,,
    \end{split}
\label{eq:spectral}
\end{equation}
where $d_i$ are functions of $C_i$ of Eq.~\eqref{eq:operator}.
We found that the final performance of both PMMs at NLO is comparable but that the PMM that uses 6 spectral LECs converges faster and more consistently.
This can be understood as fewer matrix elements need to be fit to data.
We explored bounds on the matrix elements when fitting and found that setting the bounds on each matrix element to $[-500,500]$~fm$^{-3(-5)}$ for LO (NLO) contacts ensures that the eigenvalues are well reproduced and reasonable.
We tested the effect of varying N$_{\rm dim}$ and N$_{\rm train}$ on the PMM performance at NLO and found that the percent error levels off at about $1$\% error for N$_{\rm train}\gtrsim 20$ when N$_{\rm dim} \geq 2$.
This lower bound on the error is probably due to the statistical uncertainty of the AFDMC calculations of the same order and is small compared to uncertainties due to the truncation of the chiral expansion, which is of the order of 10\% at saturation density~\cite{Tews:2024owl}.
Therefore, we do not expect this emulator uncertainty to affect the conclusions in this Letter.
Results at NLO are given in the Supplemental Material~\cite{supp}.

\begin{figure*}[t]
    \centering
    \includegraphics[width=0.61\columnwidth]{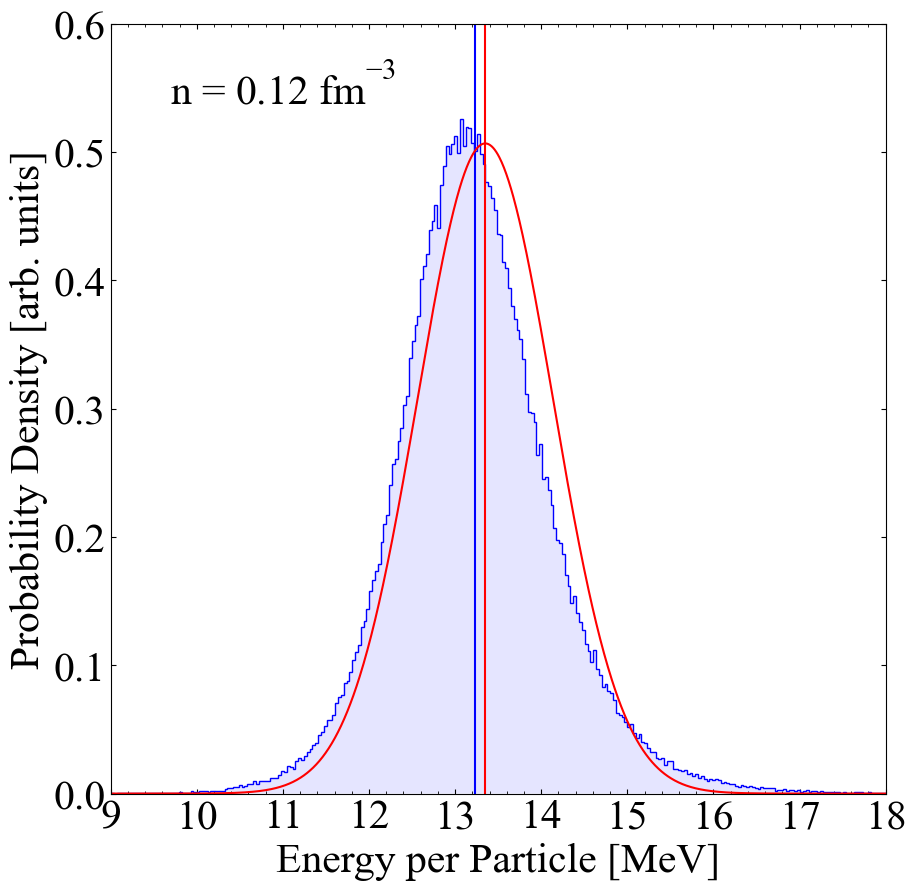} \hfil
    \includegraphics[width=0.62\columnwidth]{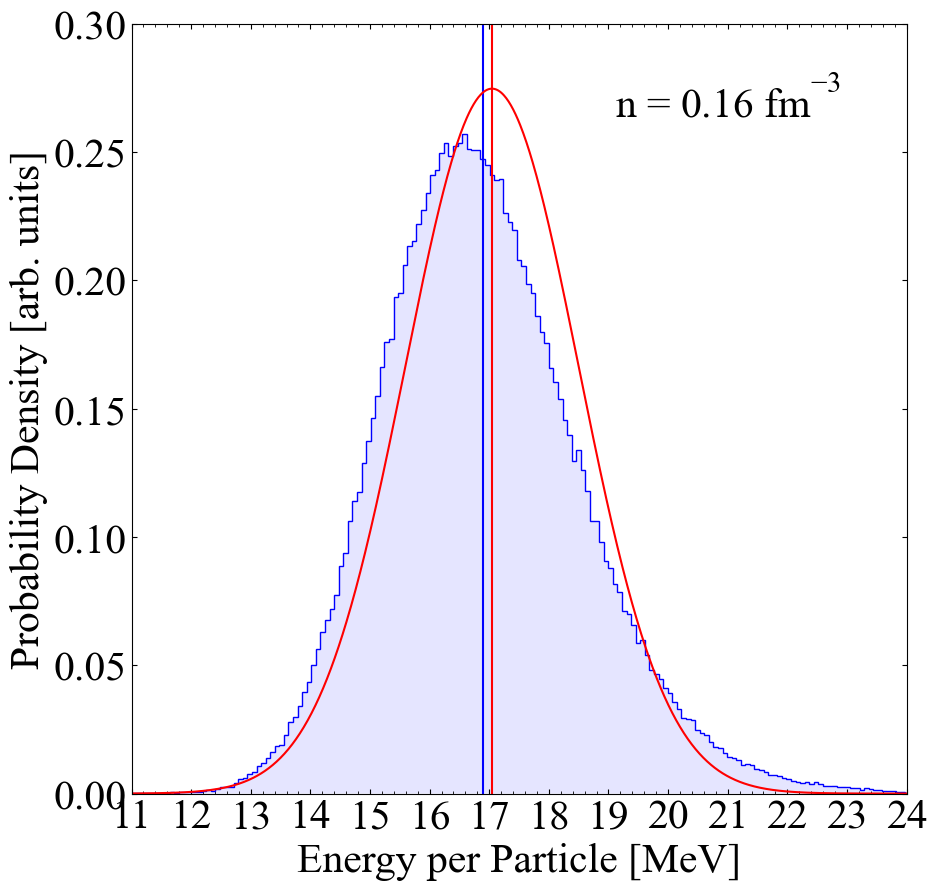}\hfil
    \includegraphics[width=0.62\columnwidth]{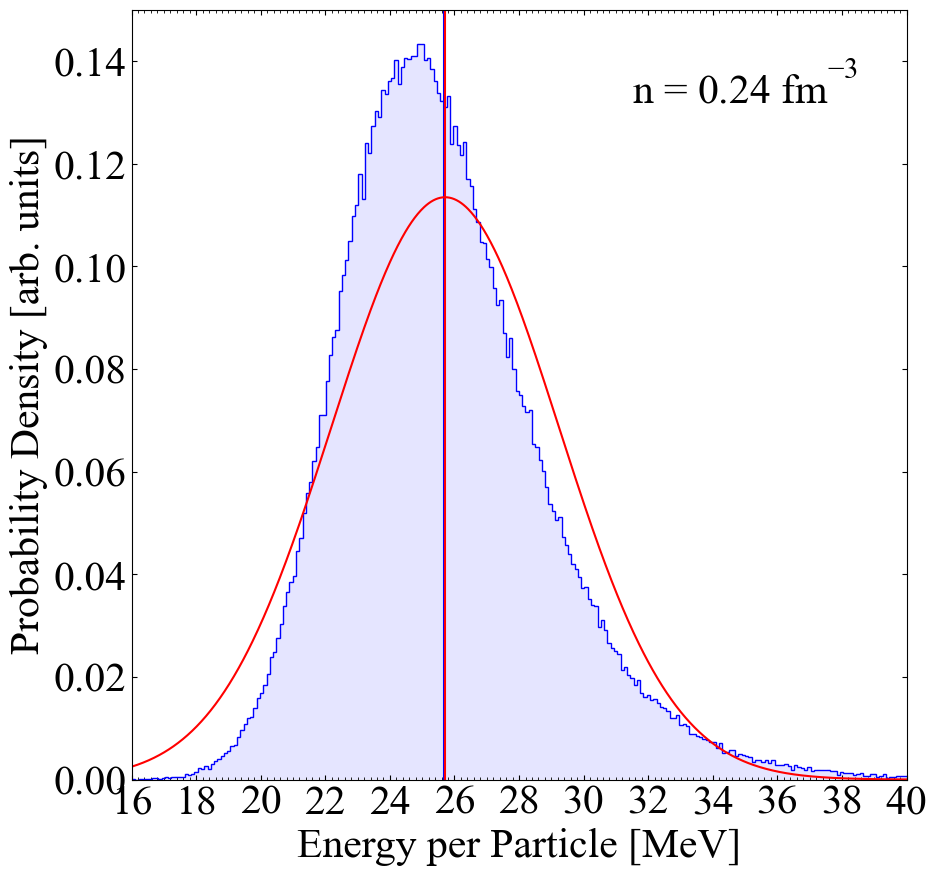} \\
    \includegraphics[width=0.62\columnwidth]{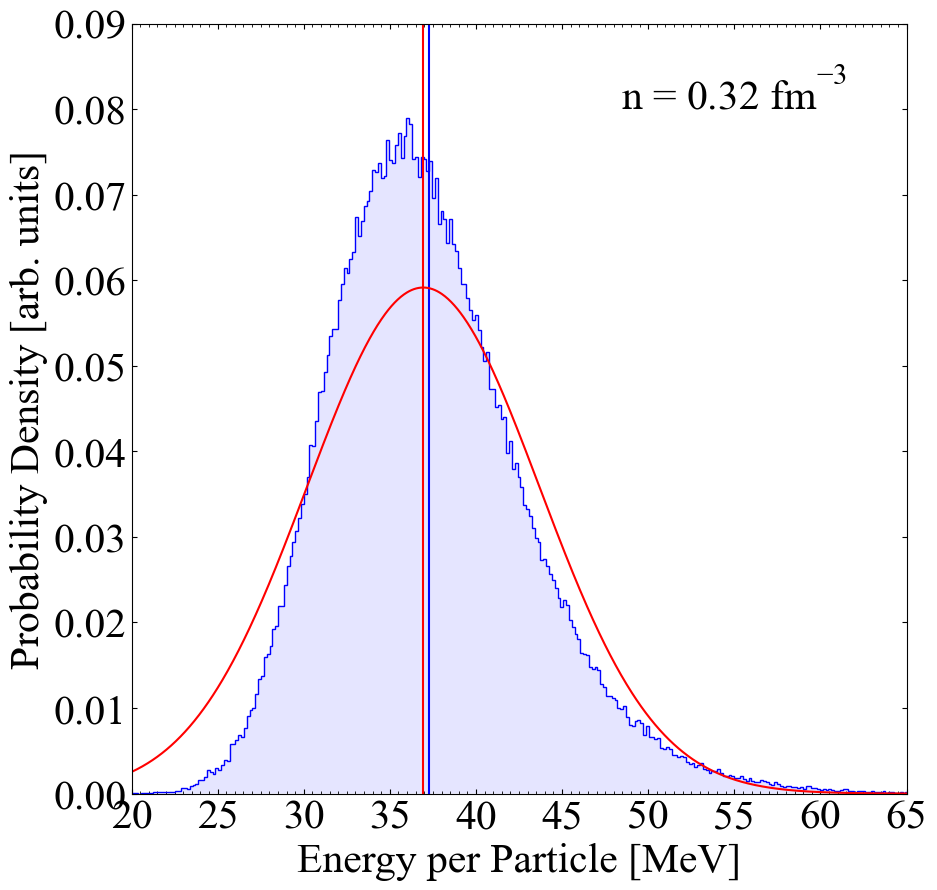}\hfil
    \includegraphics[width=0.63\columnwidth]{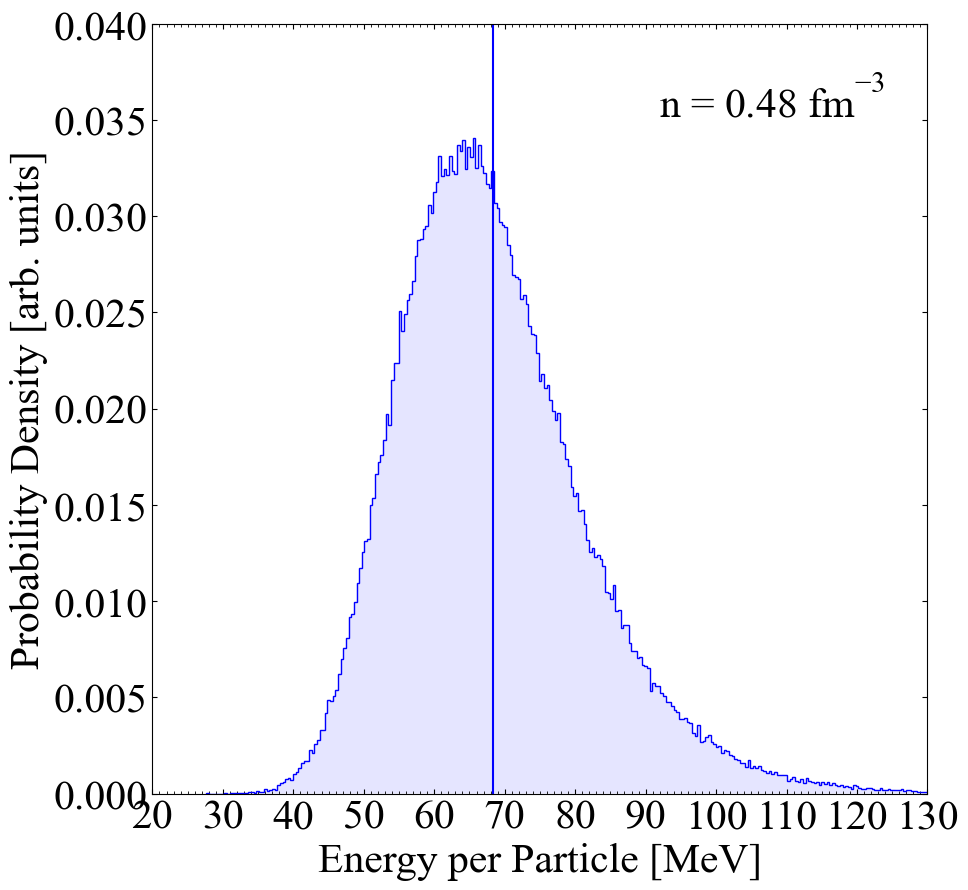}\hfil
    \includegraphics[width=0.62\columnwidth]{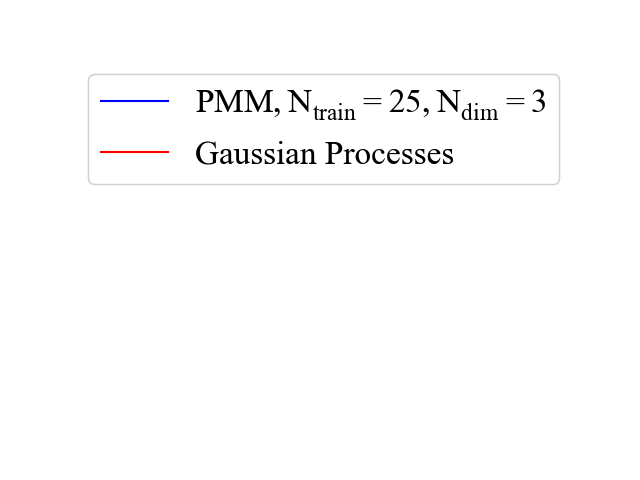}
    \caption{Probability density functions (PDFs) for the neutron-matter energy per particle at 5 different densities when propagating full posterior distributions for all LECs.
    Where available, we compare the PDFs to the previous uncertainty estimation method using Gaussian Processes~\cite{Drischler:2020hwi,Tews:2024owl}.
    For both distributions, we specify the mean by a vertical line.
    }
    \label{fig:EN_histograms}
\end{figure*}

With the fitting protocol described above and using a PMM for spectral LECs [see Eq.~\eqref{eq:spectral}], we now turn to N$^2$LO.
The calculations at this order were done with a larger number of neutrons and AFDMC walkers for 30 LEC sets.
We split these 30 calculations into N$_{\rm train}$ training data and use the remaining points for validation.
At N$^2$LO, we find that the percent errors again level off at around $1$\% for N$_{\rm train}\gtrsim 15$ for N$_{\rm dim}=1-4$, as shown in Figure~\ref{fig:error_training_lsq_n2lo} for nuclear saturation density (results are similar at other densities).
We found that the PMM for N$_{\rm dim}=5$ does not produce a good fit for a small number of training points. 
Because the quality of the PMM does not considerably improve for N$_{\rm dim}>2$, in the following we use PMMs for 25 training points and N$_{\rm dim}=2$ at each density.
We tested these choices when propagating 250,000 LEC samples and found only negligible differences in the predicted distributions of the energy per particle for larger number of training data or N$_{\rm dim}>2$ except at $3n_{\rm sat}$, where more training data led to the appearance of a level crossing.
However, this density is likely above the range of validity of chiral EFT even for high-cutoff interactions, but we include it here for future checks of the range of validity of chiral EFT.
We note that the PMMs are trained at each density independently.
We also stress that while N$_{\rm dim}=1$ leads to a surprisingly good validation error in Fig.~\ref{fig:error_training_lsq_n2lo}, this is generally not the case and depends on the order of the calculations as well as on the randomly chosen training and validation data sets.
In the Supplemental Material~\cite{supp}, we show the case at NLO where N$_{\rm dim}=1$ leads to a much worse validation error.
However, this is indicating that at N$^2$LO the dependence of the energy on the LECs is approximately linear.

{\it EOS of PNM---}Using the PMMs as specified above, we propagate 250,000 samples from the full LEC posteriors at N$^2$LO to distributions of energies per particle in PNM at $0.75n_{\rm sat}$, $n_{\rm sat}$, $1.5n_{\rm sat}$, $2n_{\rm sat}$, and $3n_{\rm sat}$.
For each LEC set, we interpolate between these 5 density points using the simple function proposed in Ref.~\cite{Gandolfi:2011xu} and use this to compute an error band in the full density range of $1-3 n_{\rm sat}$.
Our main result is shown in Fig.~\ref{fig:EOS_distribution} up to $2n_{\rm sat}$, where we compare to previous results using a Gaussian process (GP) to estimate the truncation error.
We find that the uncertainty bands are comparable but that the GP approach leads to a slightly larger 68\% confidence interval.
Furthermore, we find that the error is not symmetric.

To illustrate this point, we show histograms of the energy per particle at the 5 training densities in Figure~\ref{fig:EN_histograms}.
We find that the full distribution of energies is mostly Gaussian at lower densities but with a slight skew toward higher energies.
At higher densities, the skewness in the distribution increases.
We note that the LEC distributions resemble skewed Gaussians~\cite{Somasundaram:2023sup}.
Hence, our results show that the shape of the distribution is not conserved when propagating errors from LECs to PNM.

While the predicted EOS does not include an explicit model for truncation uncertainties, we stress that these uncertainties are absorbed in part by the LEC distributions.
We acknowledge that this approach is not fully comprehensive, especially at lower orders in the EFT. 
In particular, we estimate these uncertainties in two-nucleon scattering and absorb them only in short-range LECs while keeping the pion-exchange terms fixed.
Furthermore, our approach does not know about the structure of higher-order contributions and, therefore, fails if this structure can only be poorly absorbed by the truncated interaction.
This is especially true at LO (see Ref.~\cite{Somasundaram:2023sup}), but improves at higher orders, where corrections can be expected to be small; see Fig.~\ref{fig:EOS_distribution}.
Other approaches~\cite{Epelbaum:2014efa,Drischler:2020hwi} rely on different assumptions about the structure of these omitted terms and estimate errors based on expansion coefficients that are only informed by the truncated interactions. 
Similarly, these estimates are also problematic at lower orders in the EFT~\cite{Epelbaum:2014efa,Drischler:2020hwi}.
A strength of our approach is that it naturally enables us to account for correlated uncertainties across different nuclear systems because uncertainties are estimated from distributions of the Hamiltonian directly and not system by system. 
While no approach is without limitations, we find that different estimates agree at N$^2$LO, supporting the validity of our framework.  

\begin{figure}[t]
    \centering    
    \includegraphics[width=0.8\columnwidth]{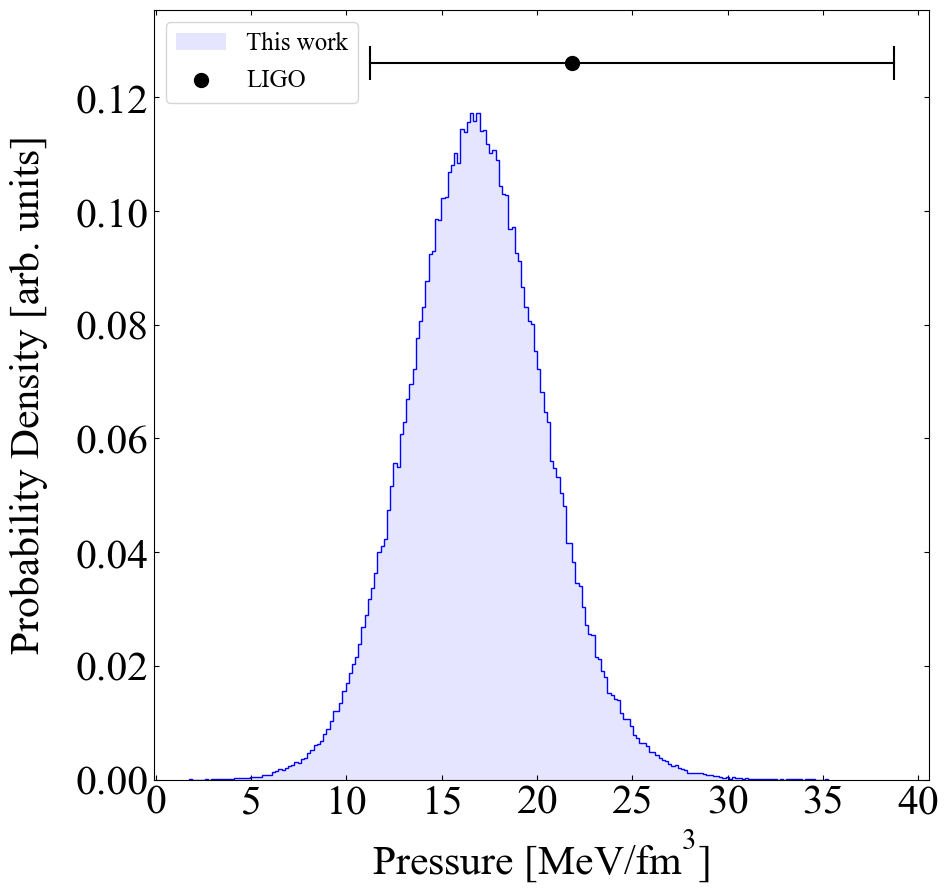}
    \caption{
    PDFs for the pressure in neutron-star matter at twice saturation density when propagating full posterior distributions for all LECs. 
    We also show the result of the LIGO collaboration of Ref.~\cite{LIGOScientific:2018cki}.
    }
    \label{fig:pressure_2nsat}
\end{figure}

Finally, we study the impact of our framework on neutron stars, and compute the pressure in $\beta$-equilibrium at twice nuclear saturation density, $p_{\beta}(2n_{\rm sat})$.
Our error estimates for this quantity are obtained from our PNM results by using the metamodel of Refs.~\cite{Margueron:2017eqc,Margueron:2017lup} to account for the small proton and lepton fractions present in neutron stars.
We extract a pressure of $p_{\beta}(2n_{\rm sat})=16.8^{+6.0}_{-5.7}$~MeV/fm$^3$ at 90\% confidence level and show the resulting histogram in Fig.~\ref{fig:pressure_2nsat}.
This is in excellent agreement with the determination from gravitational-wave data of neutron-star merger GW170817 by the LIGO collaboration~\cite{LIGOScientific:2018cki} of $p_{\beta}(2n_{\rm sat})=21.8^{+16.9}_{-10.6}$~MeV/fm$^3$ at 90\% confidence level.
Our calculations demonstrate consistency between chiral interactions and gravitational-wave neutron star observations where uncertainties in the former are fully propagated to the EOS using emulators.

{\it Conclusion---}We have built and applied fast and accurate surrogate models to AFDMC for pure neutron matter.
These emulators are based on the PMM approach and allow us to propagate uncertainties from LECs directly to the EOS of PNM.
We have found that the PMM is an excellent approach to emulating AFDMC for dense matter, faithfully reproducing our high-fidelity calculations with only a small number of high-fidelity training points.
However, care is necessary when building these emulators as level crossings can appear in the final fit results.

The LEC uncertainties generated in the fit of Ref.~\cite{Somasundaram:2023sup} include an estimate of the truncation uncertainty of the chiral EFT expansion.
Using our PMM emulators, we have propagated LEC posteriors accounting for this source of uncertainty to the EOS of neutron matter at multiple densities and found distributions of the energy per particle that are skewed at each density, which the skewness increasing with density.
Our final EOS uncertainty band differs from the previous uncertainties found using GPs.
While producing the training data for the emulators cost 22.4 million CPU-hours (30-60 LEC sets depending on the order), the final results are obtained for 250,000 LEC sets in $\sim 6.5$ seconds on commodity hardware.

Our results are important for understanding astrophysical observations of neutron stars and their mergers and provide robust nuclear-theory input for analyses of multi-messenger data. 
As we will explore in future work, the use of fast emulators will allow us to sample over LECs in inferences of astrophysical data, enabling fits of LECs directly from this data~\cite{Somasundaram:2024ykk}.

\acknowledgements
We thank E. Bonilla, J.~Carlson, S.~Gandolfi, and B.~Reed for useful discussions.
This document has been approved for unlimited release, and was assigned LA-UR-24-27320. 

C.L.A. was supported by the Laboratory Directed Research and Development (LDRD) program of Los Alamos National Laboratory (LANL) under project number 20230315ER and by LANL through its Center for Space and Earth Science, which is funded by LANL’s LDRD program under project number 20240477CR.
P.G. and K.G. were supported by the National Science Foundation CSSI program under award No.~OAC-2004601 (BAND Collaboration).
R.S. acknowledges support from the Nuclear Physics from Multi-Messenger Mergers (NP3M) Focused Research Hub which is funded by the National Science Foundation under Grant Number 21-16686, and from the LDRD program of LANL under project number 20220541ECR.
I.T. was supported by the U.S. Department of Energy, Office of Science, Office of Nuclear Physics, under contract No.~DE-AC52-06NA25396, by the U.S. Department of Energy, Office of Science, Office of Advanced Scientific Computing Research, Scientific Discovery through Advanced Computing (SciDAC) NUCLEI program, and by the LDRD program of LANL under project numbers 20220541ECR and 20230315ER.

Computational resources have been provided by the Los Alamos National Laboratory Institutional Computing Program, which is supported by the U.S. Department of Energy National Nuclear Security Administration under Contract No.~89233218CNA000001, and by the National Energy Research Scientific Computing Center (NERSC), which is supported by the U.S. Department of Energy, Office of Science, under contract No.~DE-AC02-05CH11231.

\bibliography{main}


\clearpage
\onecolumngrid

\section*{Supplemental Material}

\renewcommand{\theequation}{S\arabic{equation}} 
\setcounter{equation}{0} 

\setcounter{page}{1}   
\setcounter{figure}{0}    

In the following, we will give additional details on the PMMs constructed in this work.

\subsection*{PMM for a toy model}

\begin{figure*}[h!]
    \centering
    \includegraphics[width=0.31\textwidth]{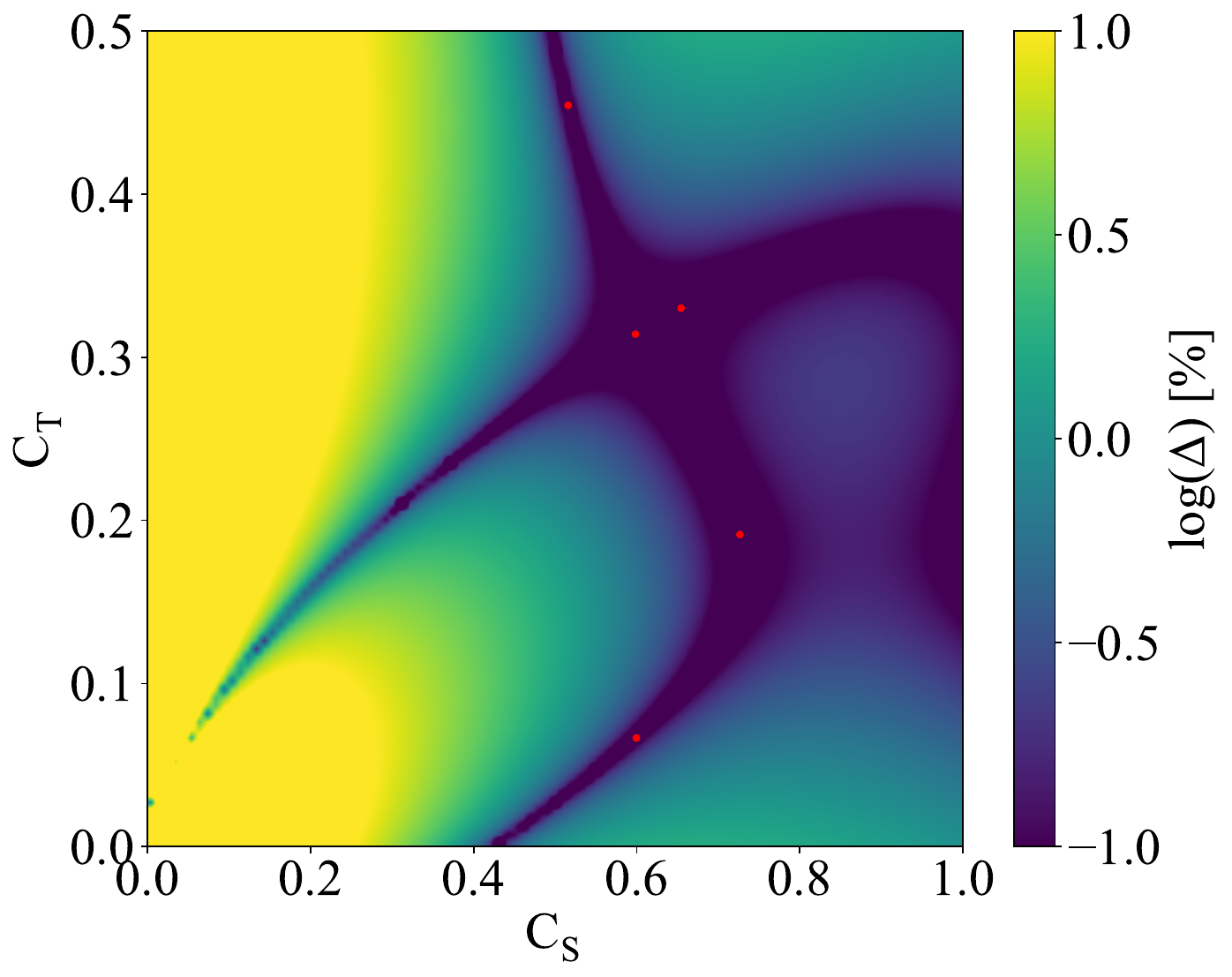}
    \quad
    \includegraphics[width=0.31\textwidth]{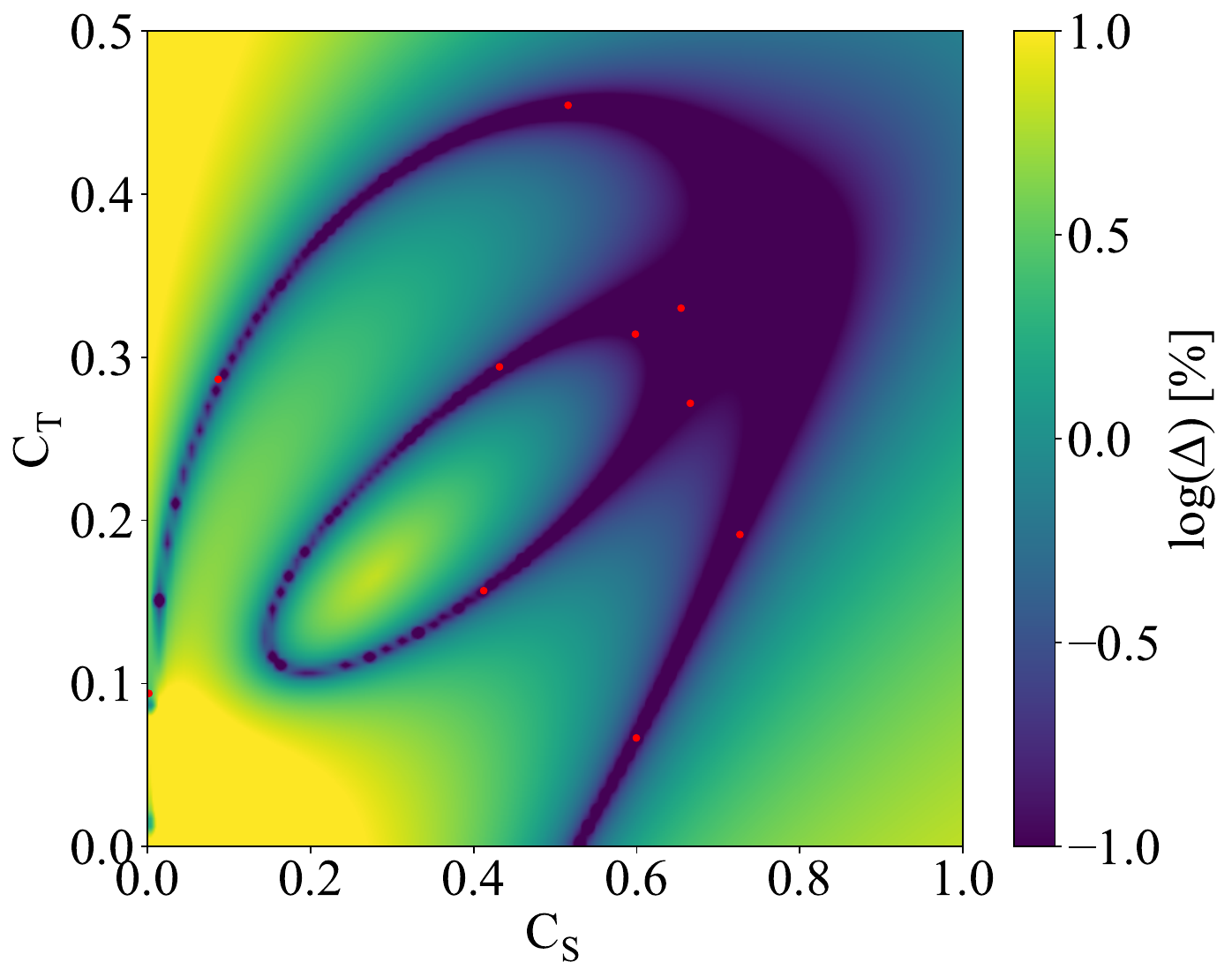}
    \quad
    \includegraphics[width=0.31\textwidth]{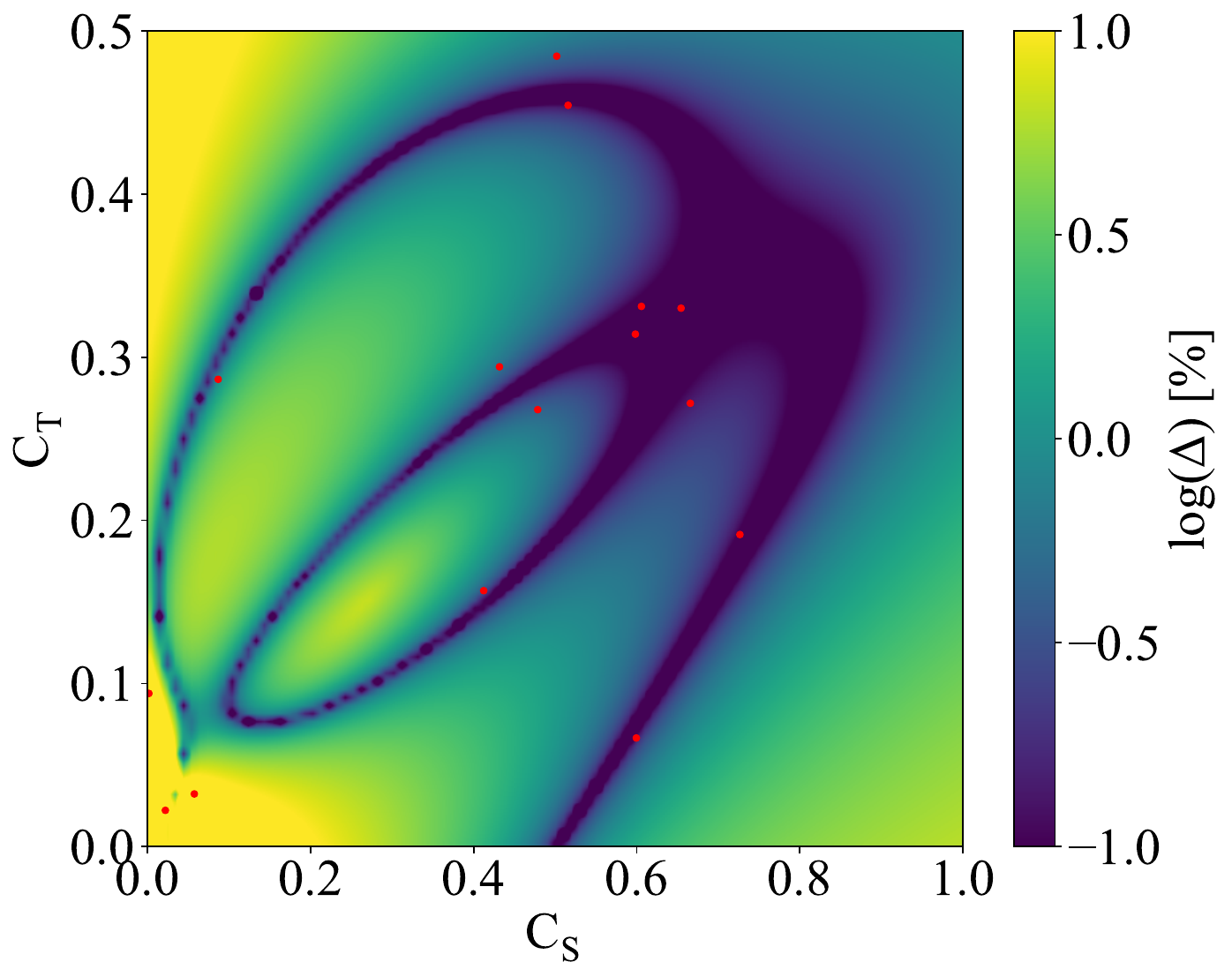}
    \quad
    \includegraphics[width=0.31\textwidth]{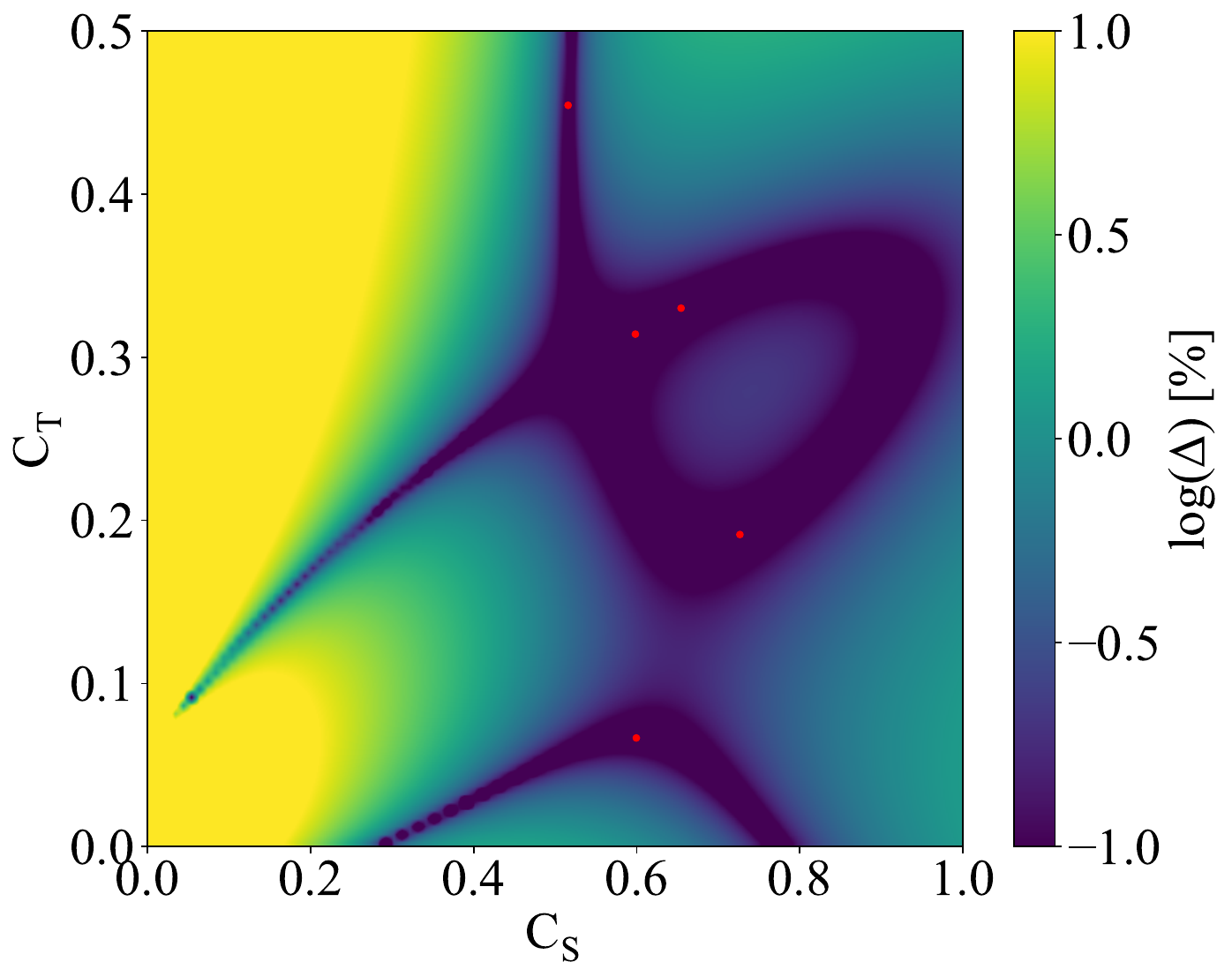}
    \quad
    \includegraphics[width=0.31\textwidth]{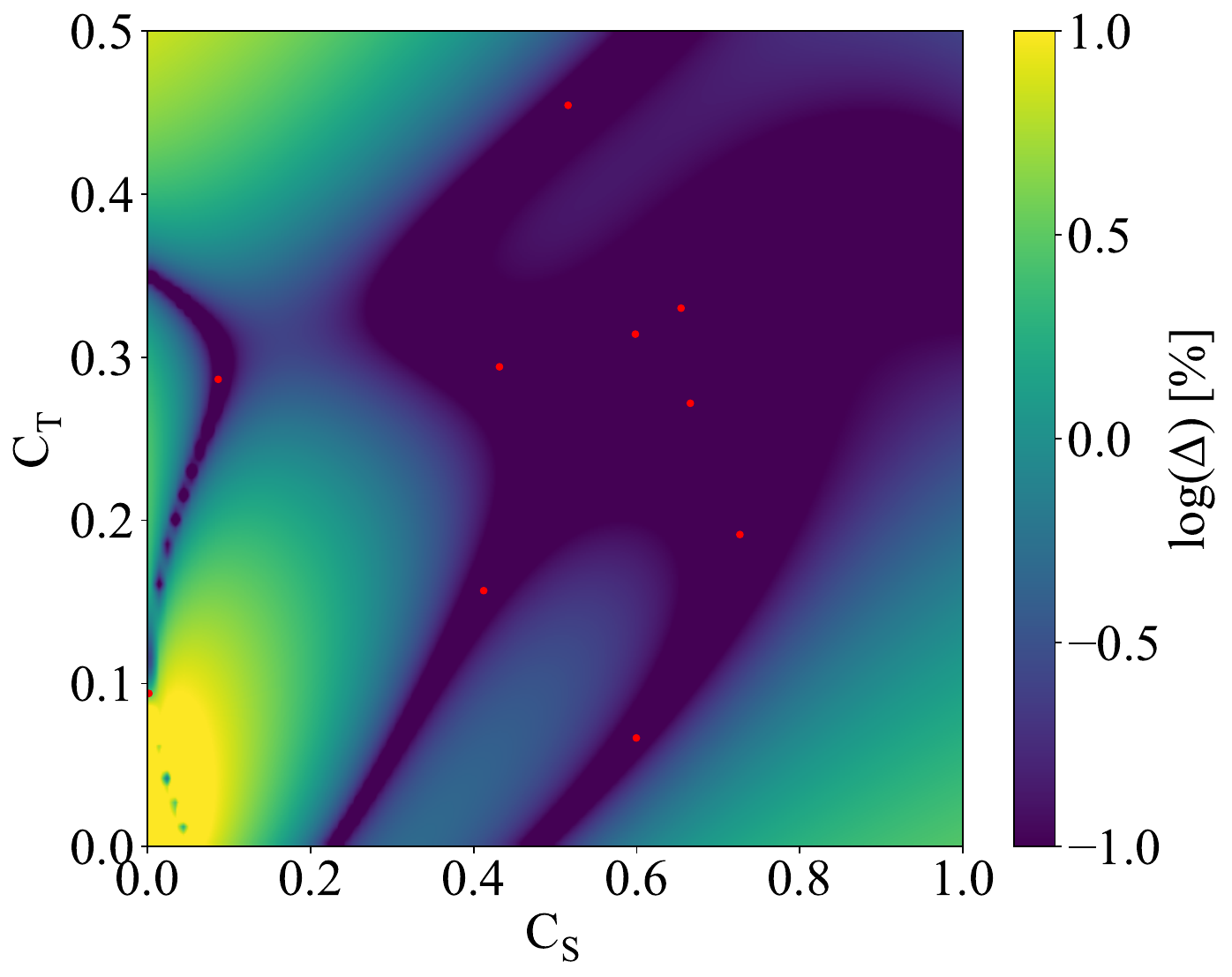}
    \quad
    \includegraphics[width=0.31\textwidth]{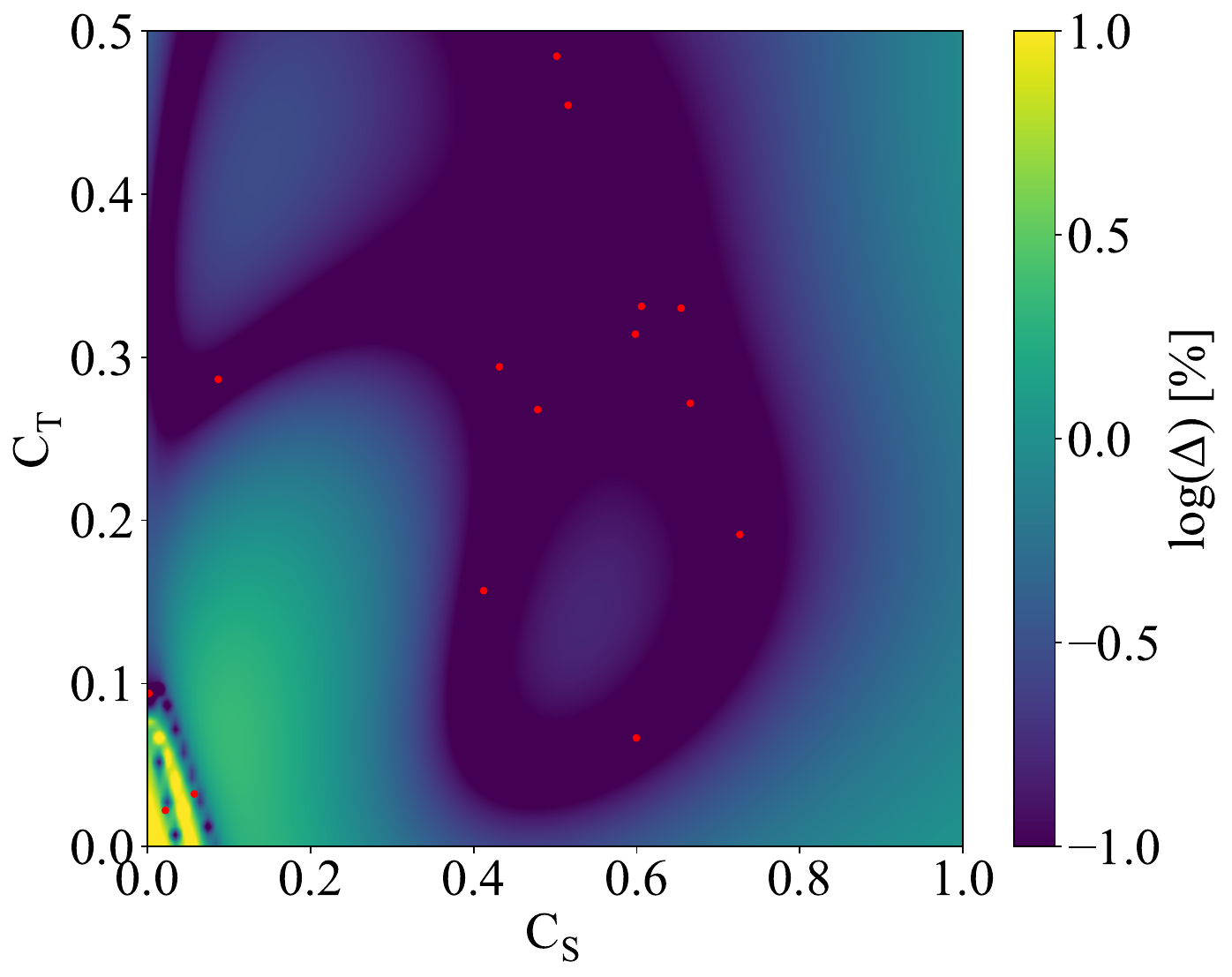}
    \quad
    \includegraphics[width=0.31\textwidth]{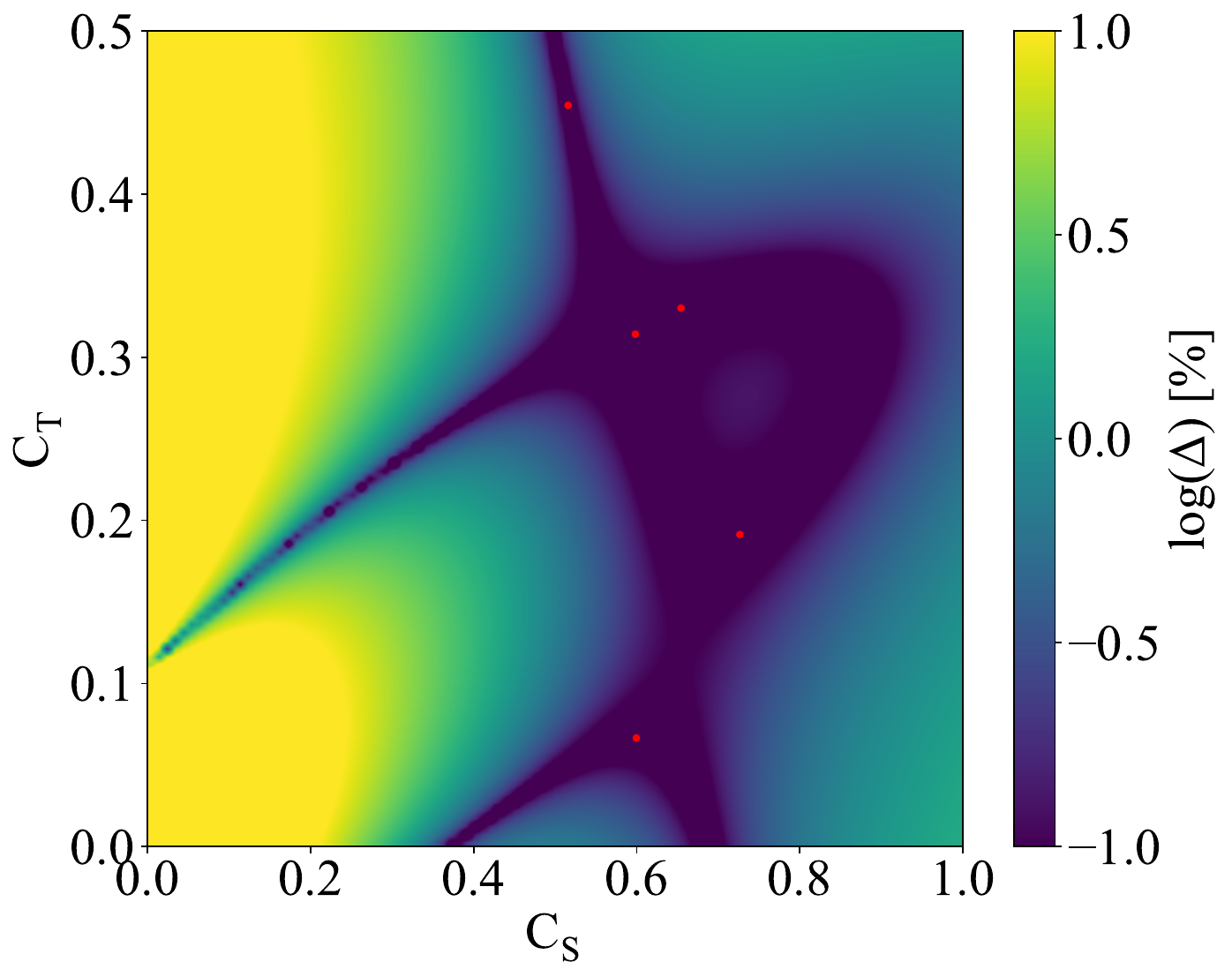}
    \quad
    \includegraphics[width=0.31\textwidth]{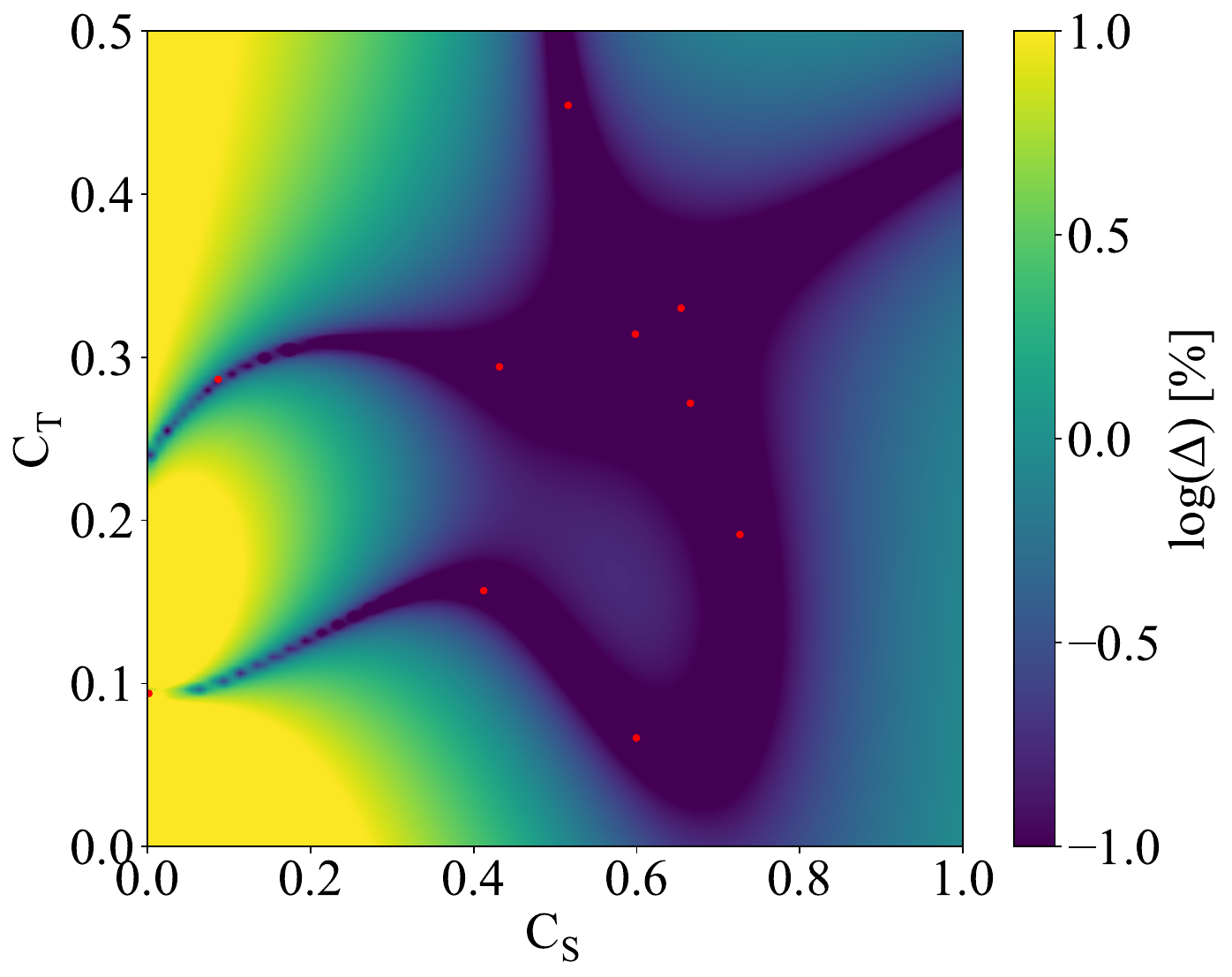}
    \quad
    \includegraphics[width=0.31\textwidth]{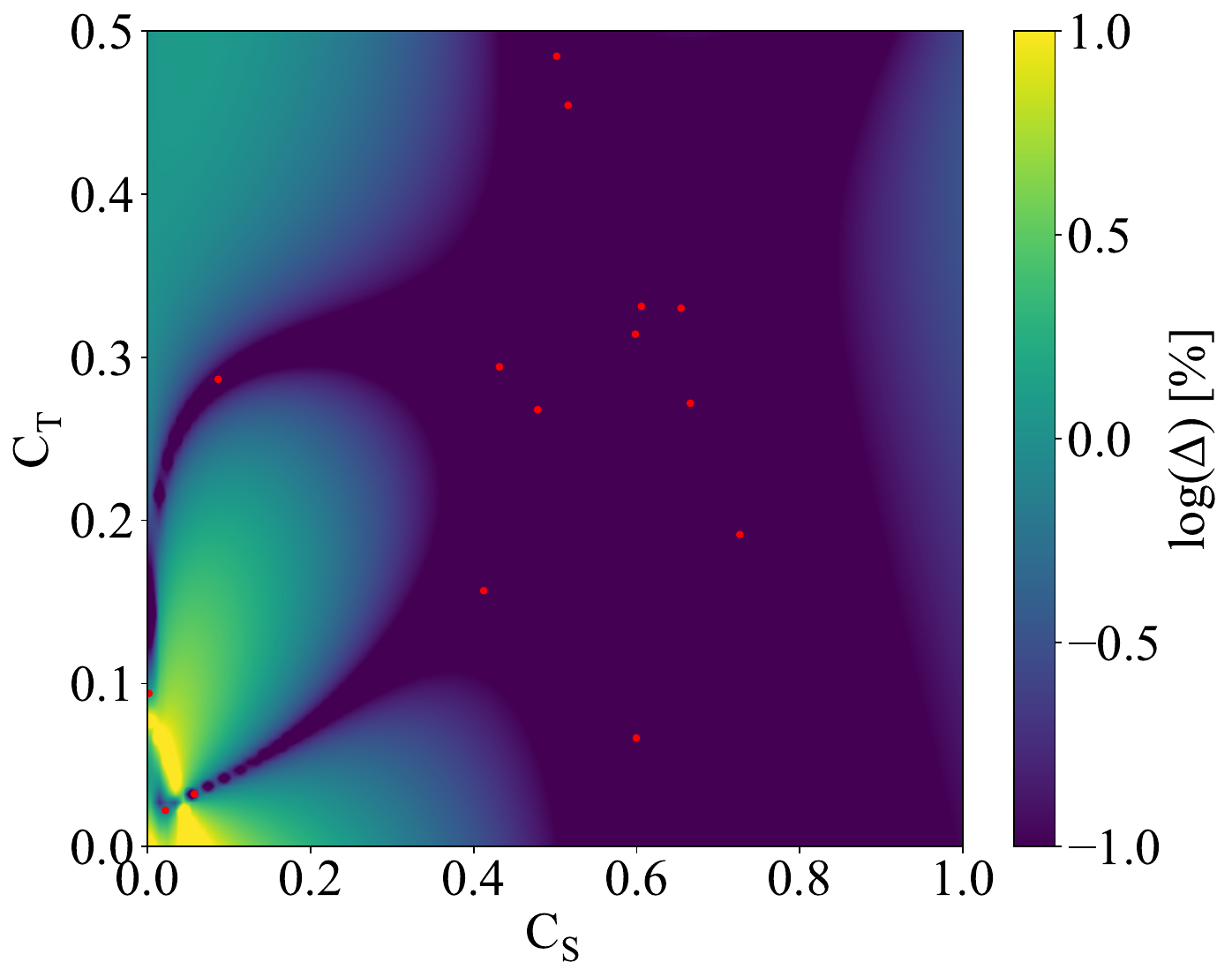}
    \quad
    \includegraphics[width=0.31\textwidth]{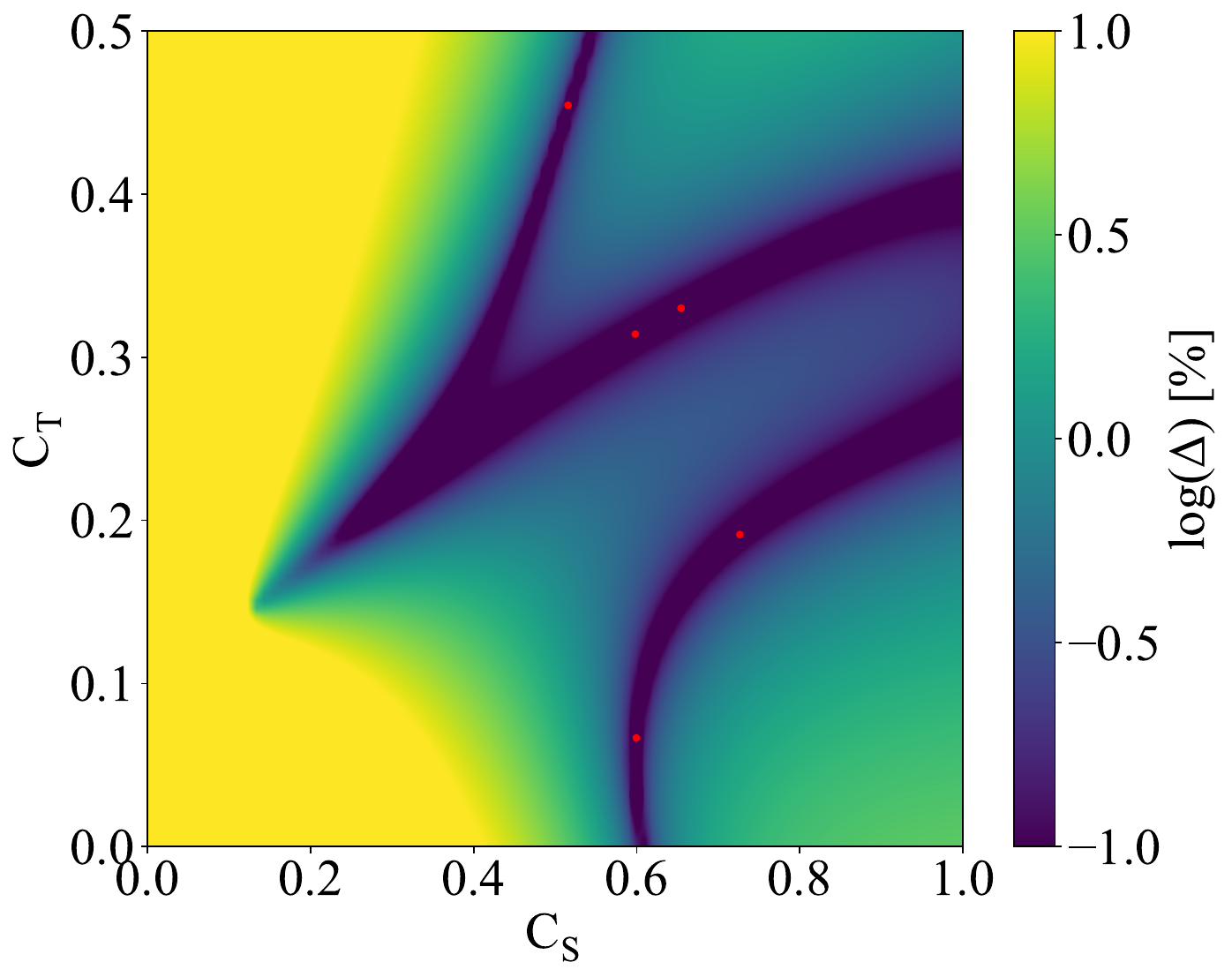}
    \quad
    \includegraphics[width=0.31\textwidth]{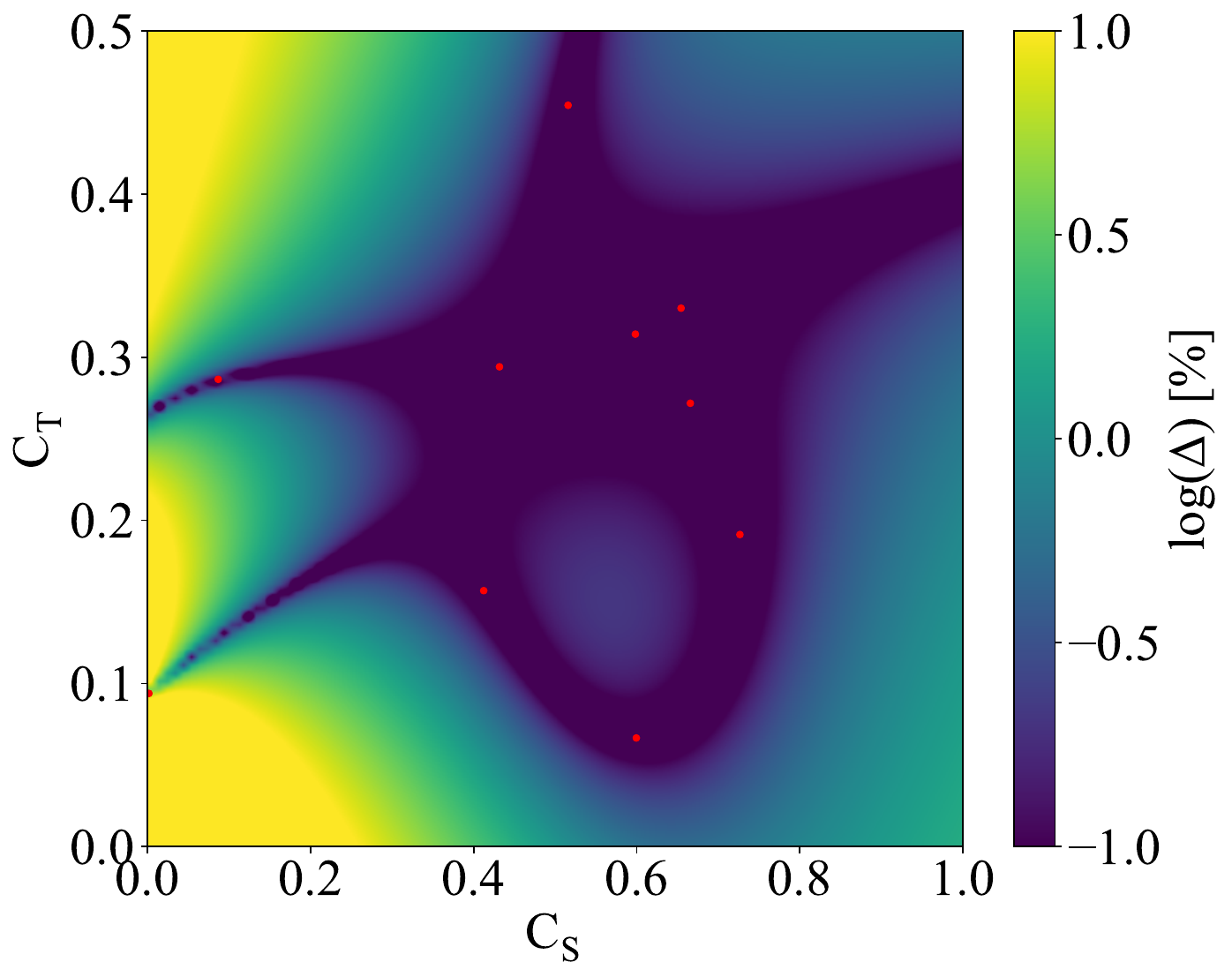}
    \quad
    \includegraphics[width=0.30\textwidth]{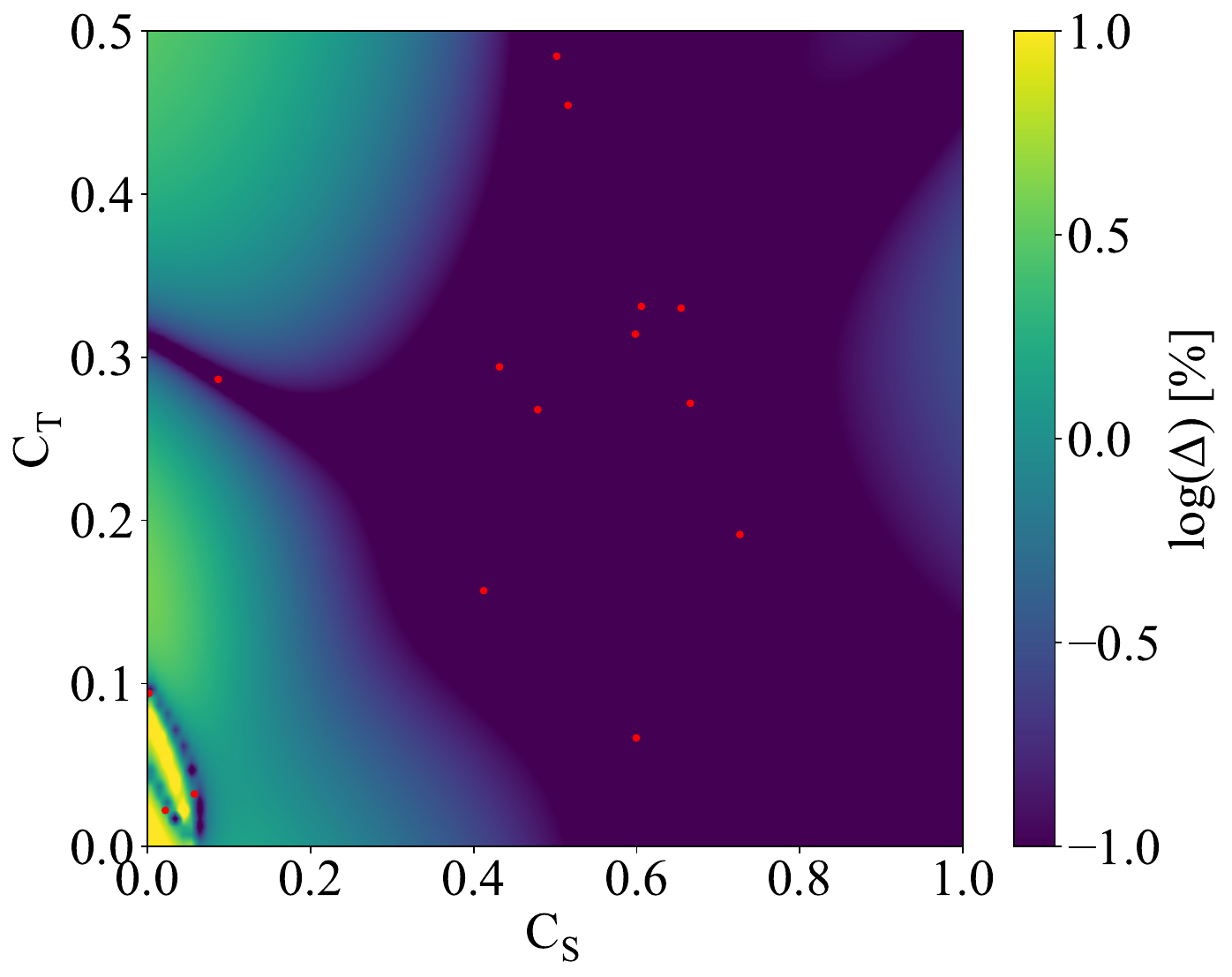}
    \caption{Accuracy for reproducing eigenvalues of a randomly generated 10-dimensional eigenvalue equation.
    We show the logarithm of the relative error between the full 10-dimensional model and PMM emulation for different numbers of training points. 
    From left to right, the PMM is trained on 5, 10, and 15 points which are shown in red. 
    Each subsequent row increases the number of dimensions in the matrices fit starting from 2-dimensional matrices for the top row and ending with 5-dimensional matrices for the bottom row. 
    }
    \label{fig:accuracies_for_toy}
\end{figure*}

To test the PMM approach, better understand the required number of training data for PMMs of different dimension, and asses how well PMMs extrapolate with scarce training data, we have set up a toy problem.     
We have randomly generated a 10-dimensional toy Hamiltonian with two parameters, similar to Eq.~\eqref{eq:PMM}, and exactly solved for its eigenvalues as function of the the two parameters $C_S$ and $C_T$.
We have then constructed PMMs with two to five dimensions and studied the relative errors of the PMM reproduction of the full 10-dimensional problem. 
We show the obtained relative errors in Fig.~\ref{fig:accuracies_for_toy} for 5, 10, and 15 training points.

We find that the PMM with two parameters generally works very well already with a small number of training data in parts of the parameter space that are much larger than the phase space covered by the training data.
In the 2-dimensional case, for which there are 8 unknown matrix elements, the PMM saturates between 5-10 training points as expected.
For higher-dimensional PMMs, though, the relative uncertainties are very small in large parts of the parameter space even if there are fewer training data than matrix elements to determine. 
In case of a 5-dimensional PMM (35 unknowns), the parameter space is well recovered even with 15 training data.
This highlights the power of the PMM as an efficient emulator for interpolation and extrapolation.

\begin{figure}[t]
    \centering
    
    \includegraphics[width=0.8\columnwidth]{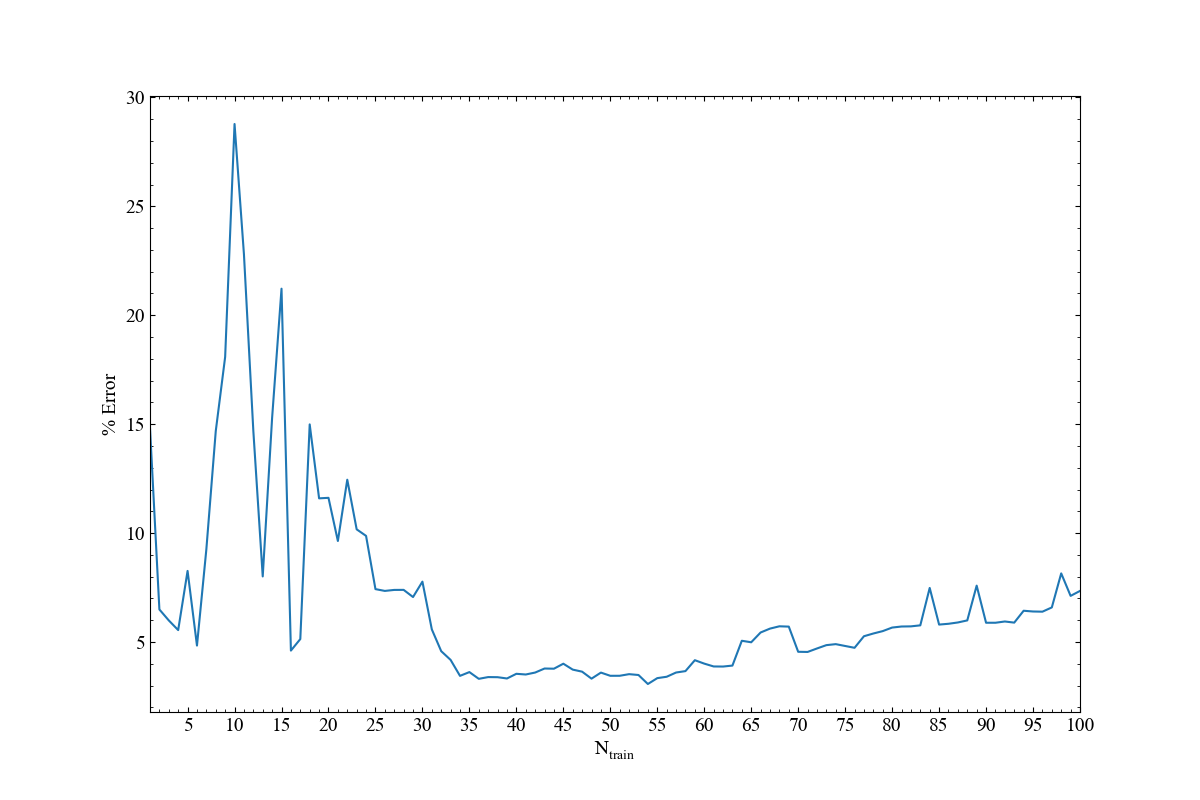}
    \caption{The evolution of the average validation error as function of N$_{\rm train} = 1-100$ for N$_{\rm dim}=2$ for a toy problem with 9 control parameters.
    }
    \label{fig:NLO_10x10_error}
\end{figure}

We have also set up a similar toy problem for a 9-parameter PMM to test the optimal number of training data to construct PMMs resembling the case at NLO and N$^2$LO with operator LECs.
We show the relative error for a two-dimensional PMM (29 unknown matrix elements) as function of number of training points N$_{\rm train}$ in Fig.~\ref{fig:NLO_10x10_error}.
We find that the error decreases and levels off at 35-55 training data, but slowly increases again for larger number of training data, likely due to overfitting.
We conclude that N$_{\rm train}$ should be similar to the number of free matrix elements, which suggests that we need $\sim 20$ training data for our PMM fits at NLO and N$^2$LO with six spectral LECs (20 matrix elements for two-dimensional PMM).

\subsection{Testing the PMM setup at LO}

\begin{figure}[h]
    \centering
   \includegraphics[trim= 1.8cm 0.5cm 2.0cm 0.5cm, clip=, width=0.7\columnwidth]{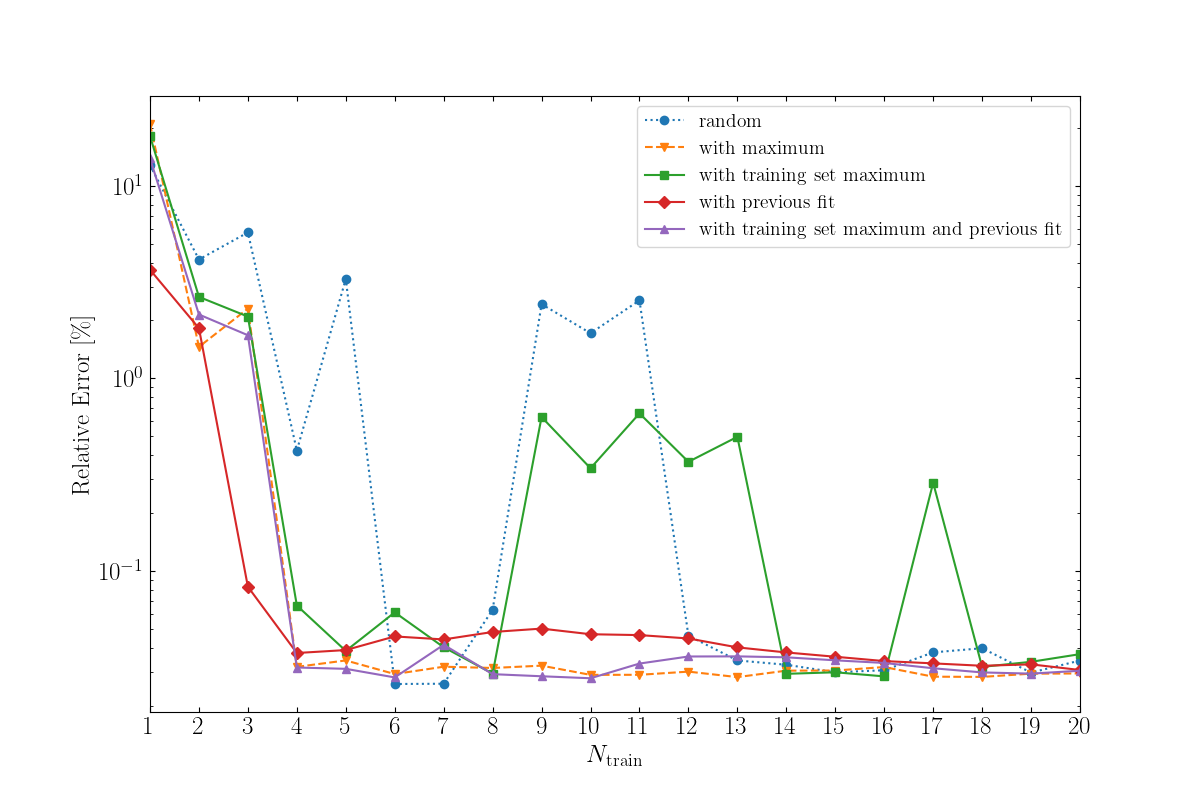}
    \caption{The relative error of a 2-dimensional PMM when reproducing the AFDMC validation energies as function of number of training points N$_{\rm train}$.
    For all lines, the AFDMC energies are first split into a training set and a validation set using different prescriptions. 
    For the blue line, the training points are randomly chosen and ordered.
    For the orange line, the training set includes the global highest-energy AFDMC result in the overall data set at first position, so all N$_{\rm train}$ include this maximum.
    For the green line, the training set is randomly chosen but with the highest AFDMC energy in the training set at first position, so all N$_{\rm train}$ include this point. 
    For the red line, the training set is randomly ordered, but the fit results for each N$_{\rm train}$ is used as a starting point for the subsequent fit with N$_{\rm train}+1$. 
    For the purple line, all N$_{\rm train}$ include the highest energy in the training set and the fit for each N$_{\rm train}$ is used as a starting point for the subsequent fit at N$_{\rm train}+1$.
    }
    \label{fig:order_comparison}
\end{figure}

\begin{figure*}[h!]
    \centering
\includegraphics[width=0.32\columnwidth]{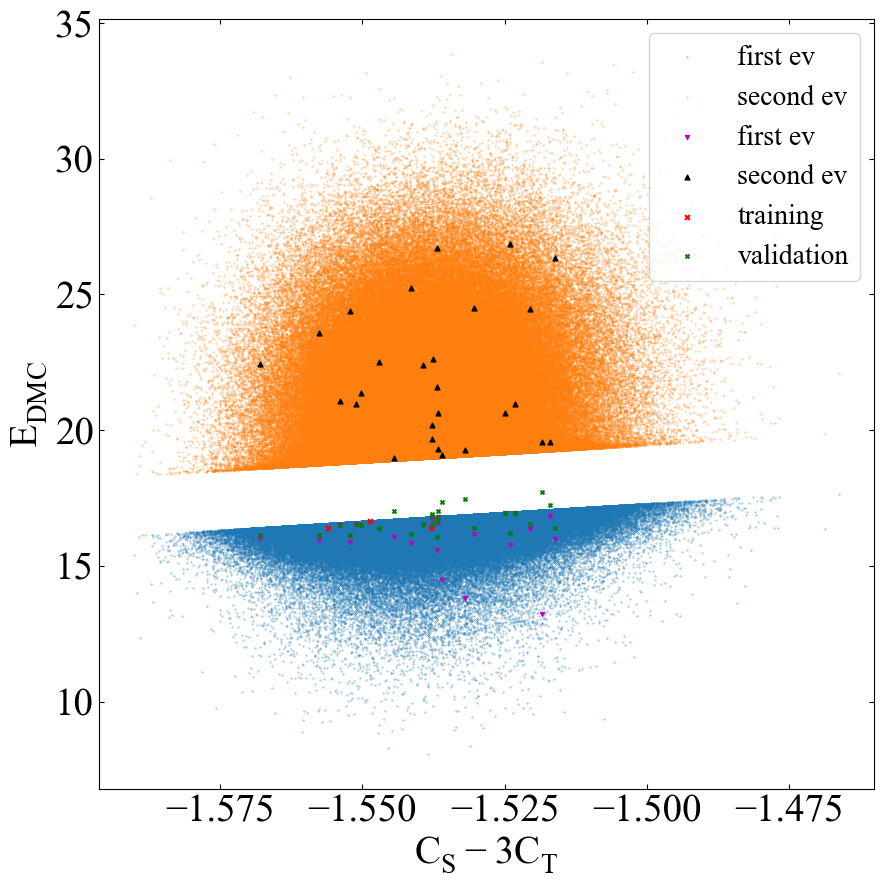}
\includegraphics[width=0.32\columnwidth]{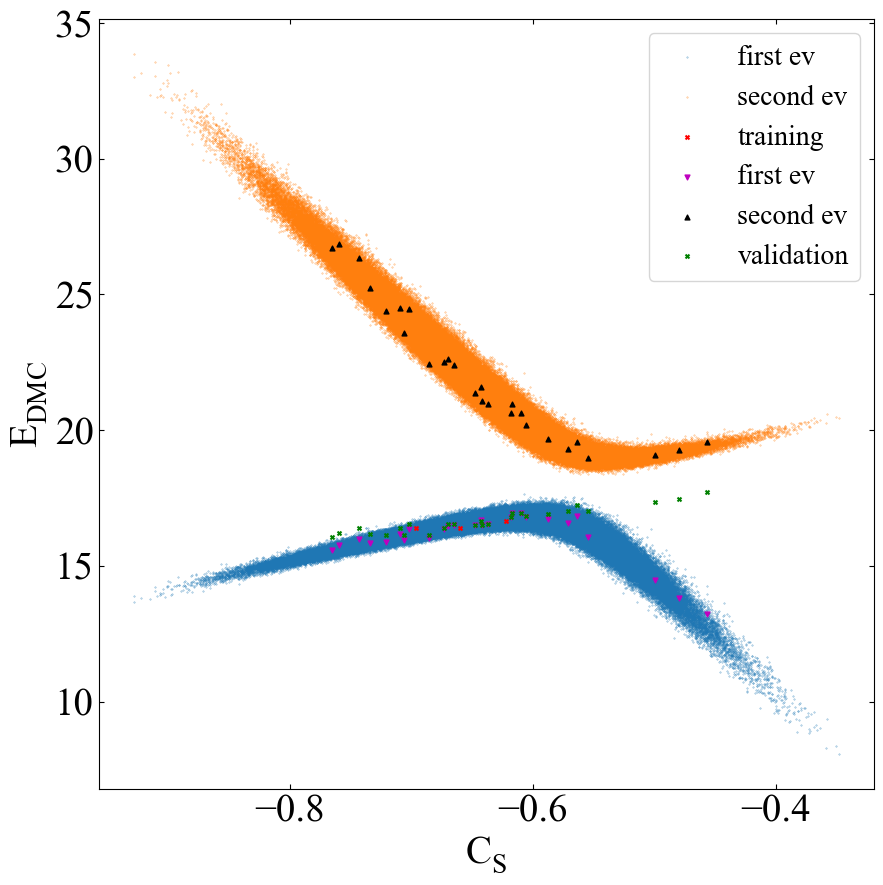}
\includegraphics[width=0.32\columnwidth]{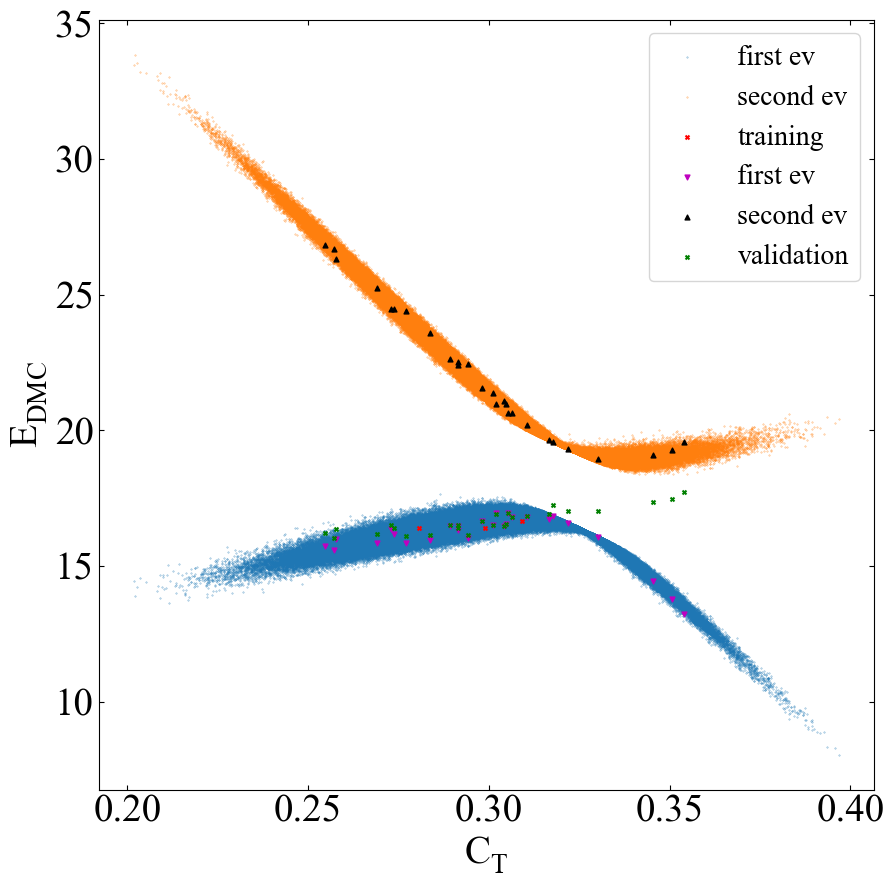}\\
\includegraphics[width=0.32\columnwidth]{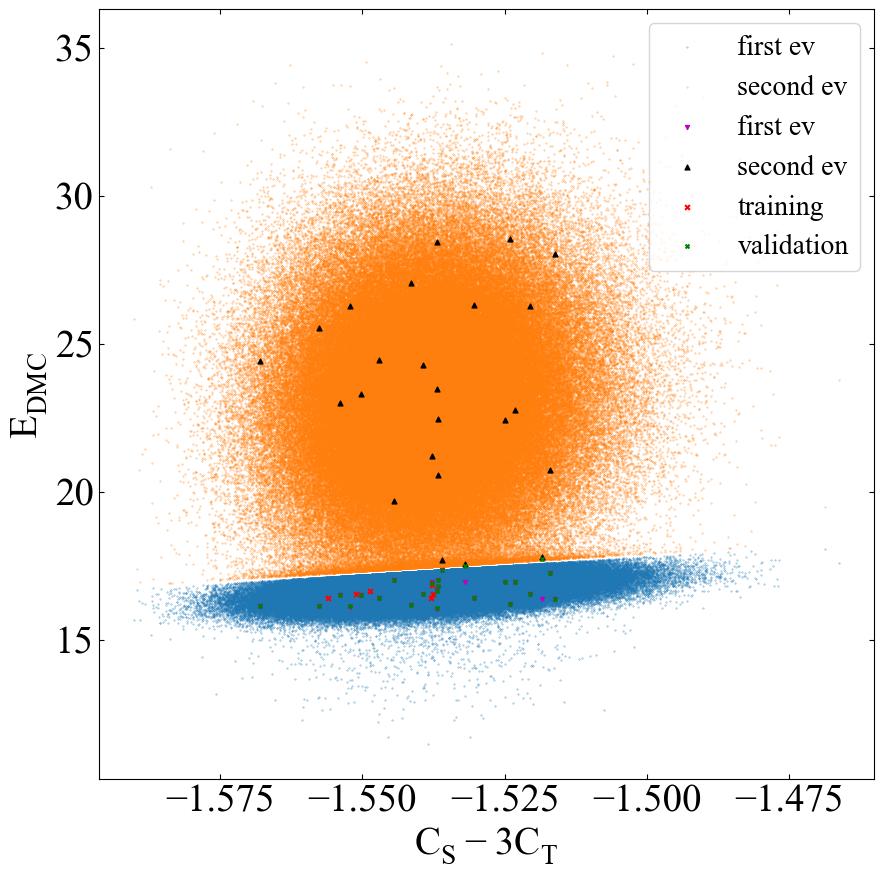}
\includegraphics[width=0.32\columnwidth]{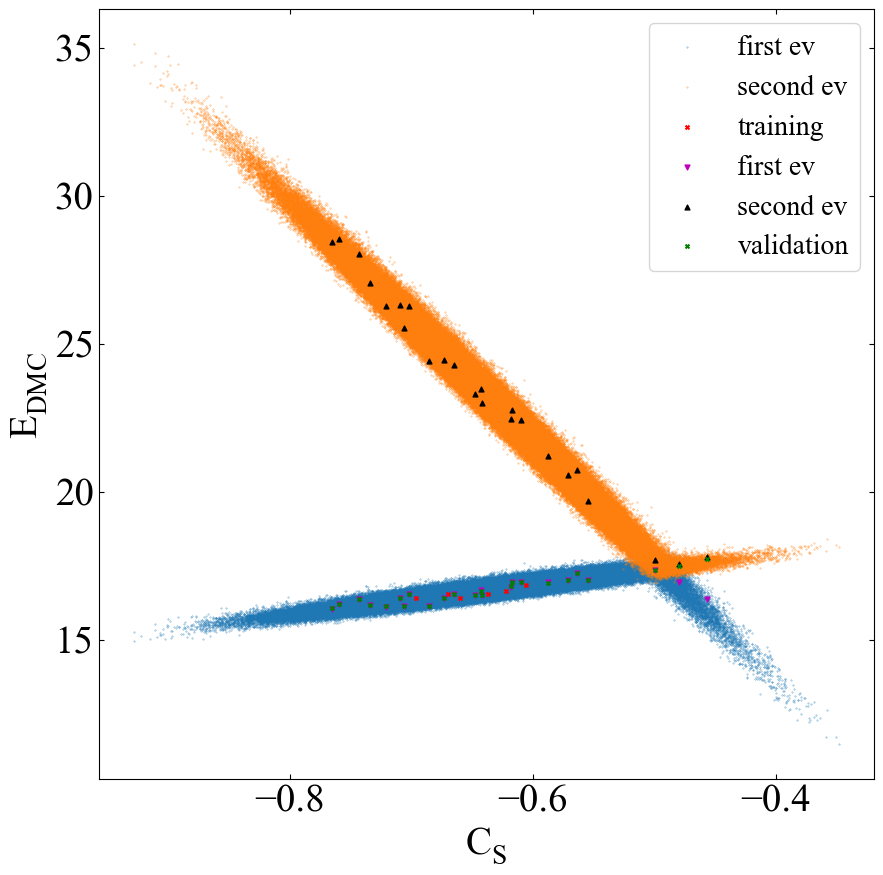}
\includegraphics[width=0.32\columnwidth]{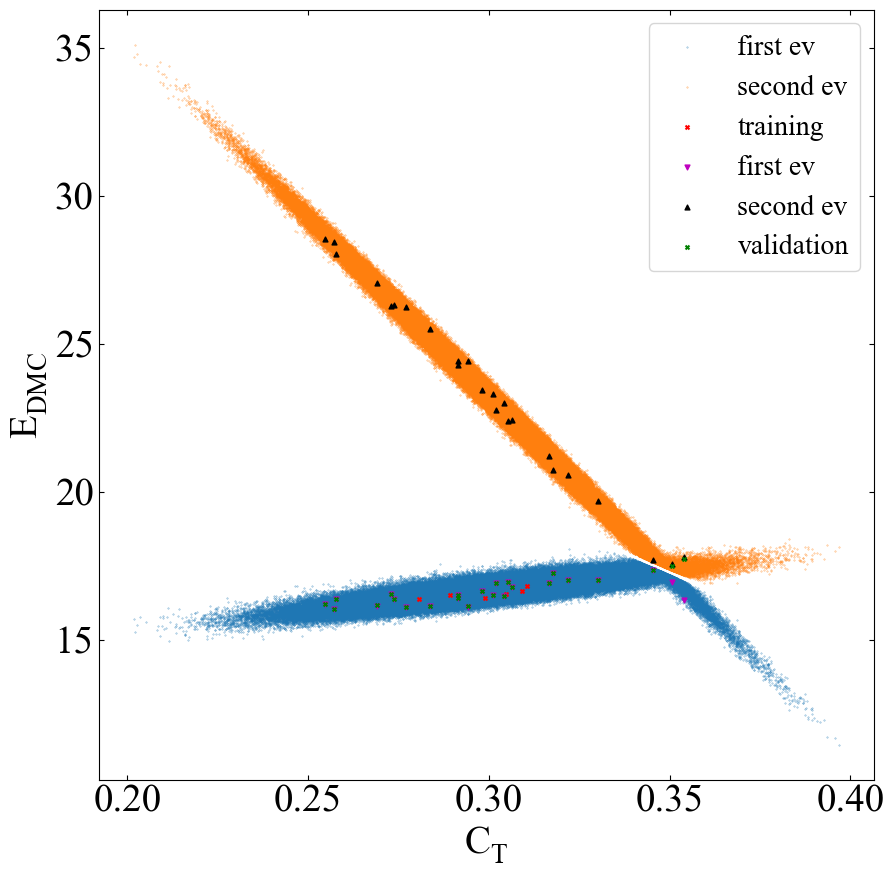}\\
\includegraphics[width=0.32\columnwidth]{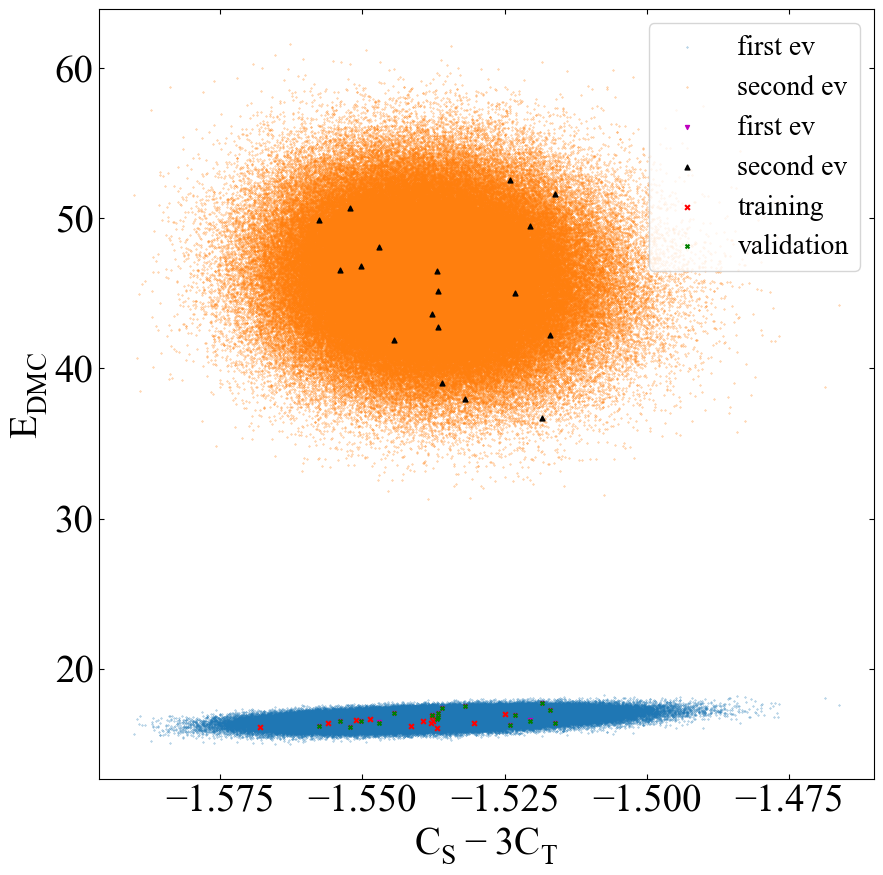}
\includegraphics[width=0.32\columnwidth]{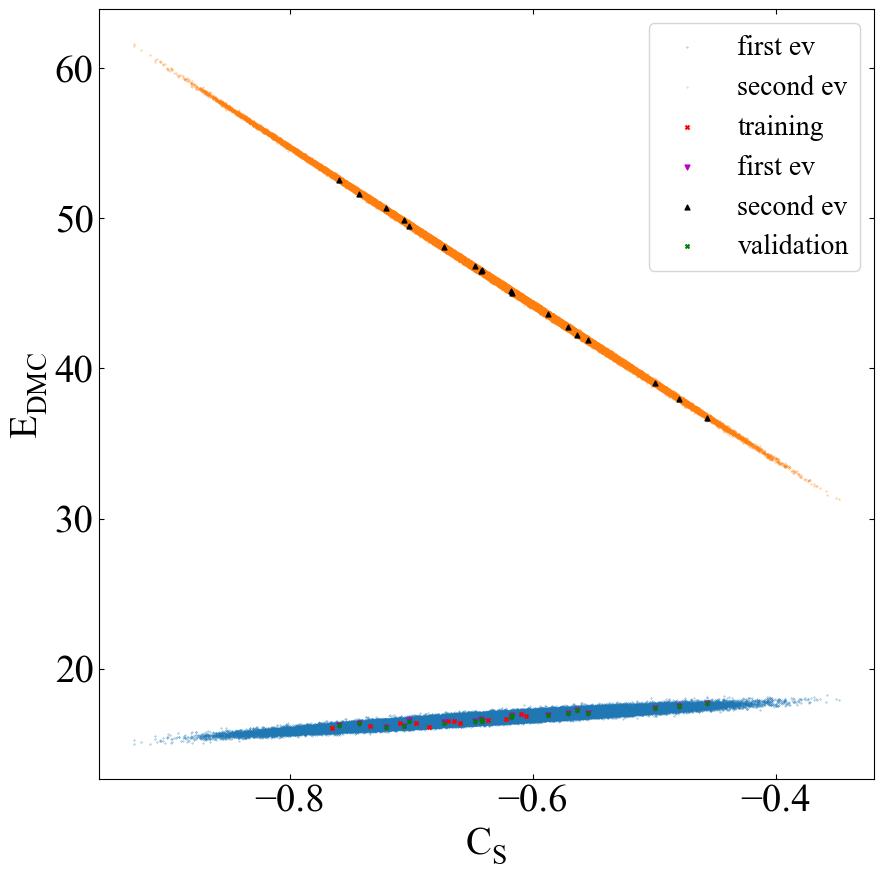}
\includegraphics[width=0.32\columnwidth]{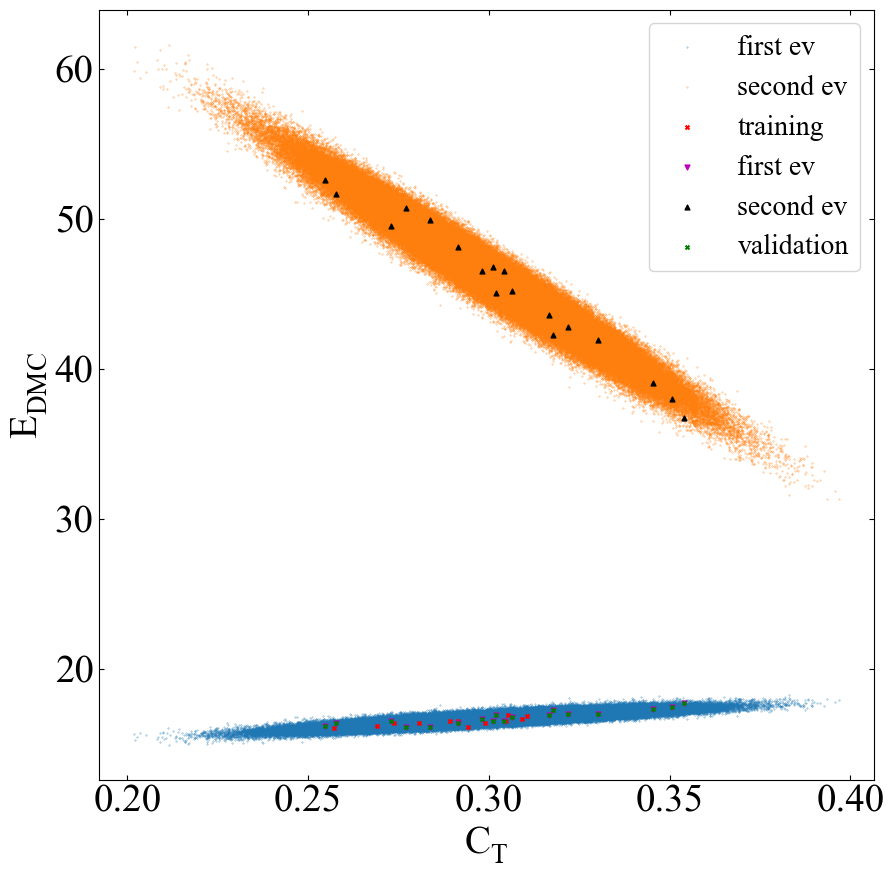}
    \caption{Correlation of AFDMC energies with $C_S-3C_T$, $C_S$, and $C_T$ for N$_{\rm train}=5$ and N$_{\rm dim}=2$. All rows are for different PMM fits and show all PMM eigenvalues if 100k LEC sets are propagated.
    The upper row is for a PMM with a level crossing and a gap in the energy spectrum.
    The middle row is for a PMM with level crossing but without a gap in the energy spectrum.
    The lower row is for a PMM without a level crossing.
    }
    \label{fig:upper_limits}
\end{figure*}

\begin{figure*}[h]
    \centering
\includegraphics[width=0.32\columnwidth]{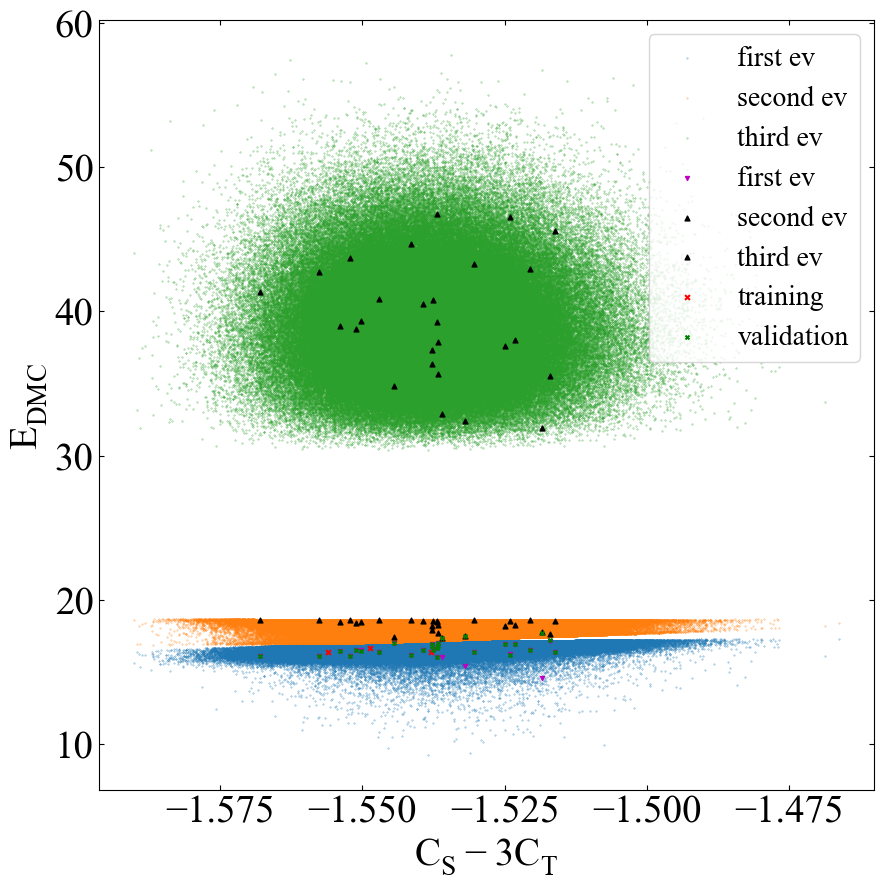}
\includegraphics[width=0.32\columnwidth]{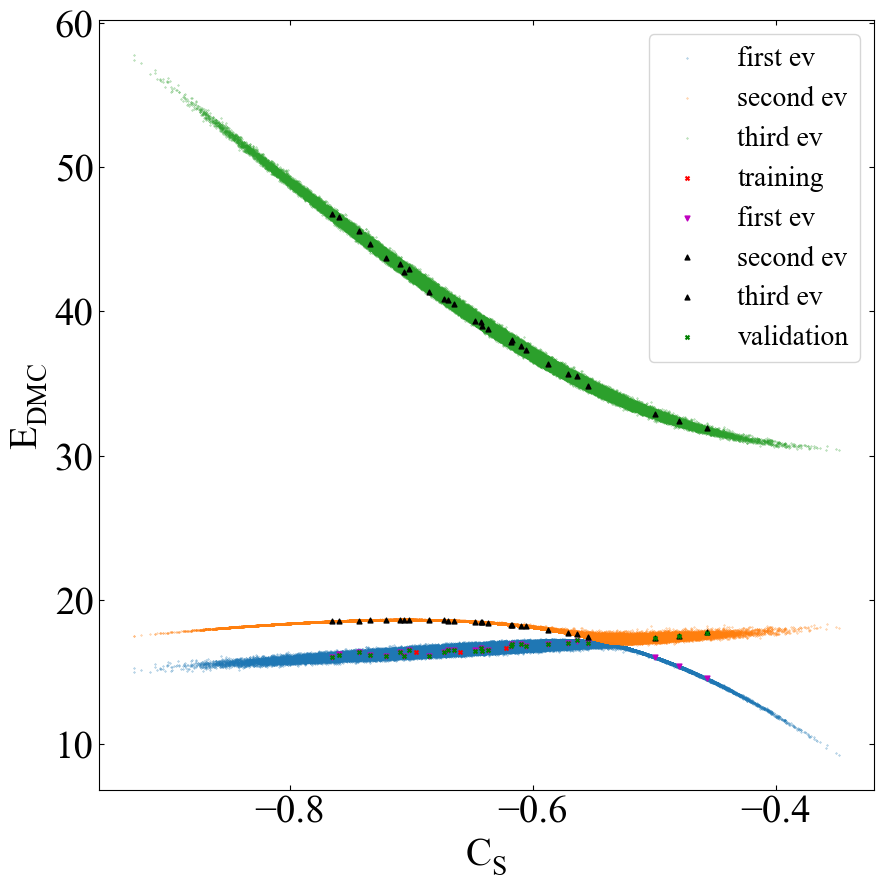}
\includegraphics[width=0.32\columnwidth]{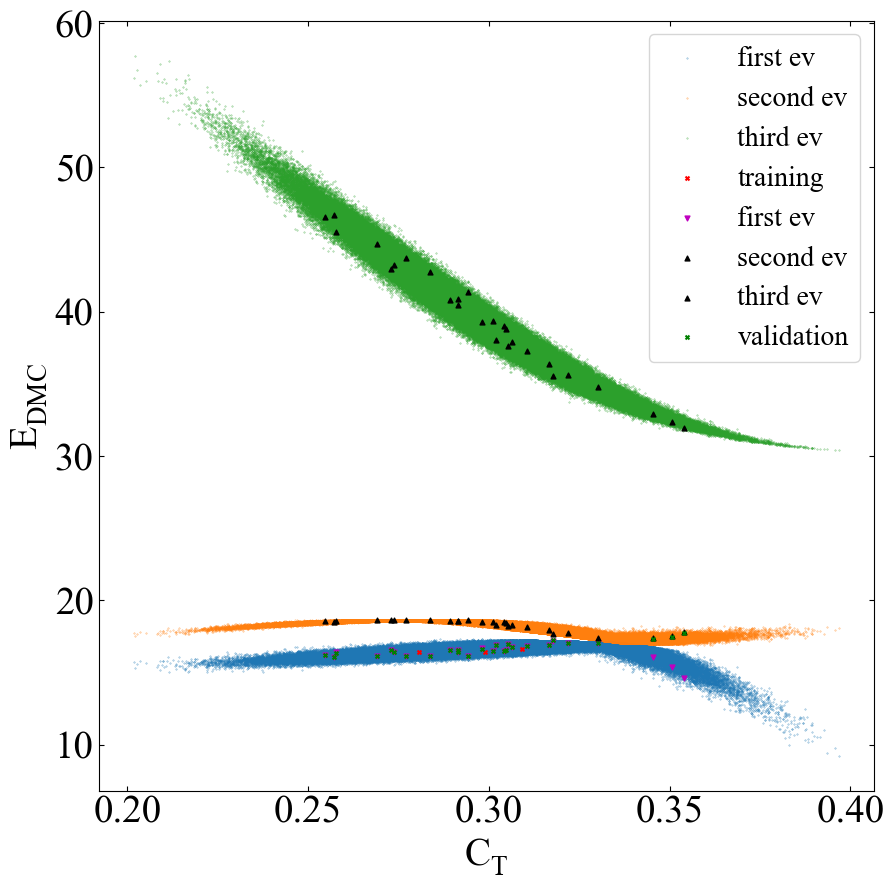}\\
\includegraphics[width=0.32\columnwidth]{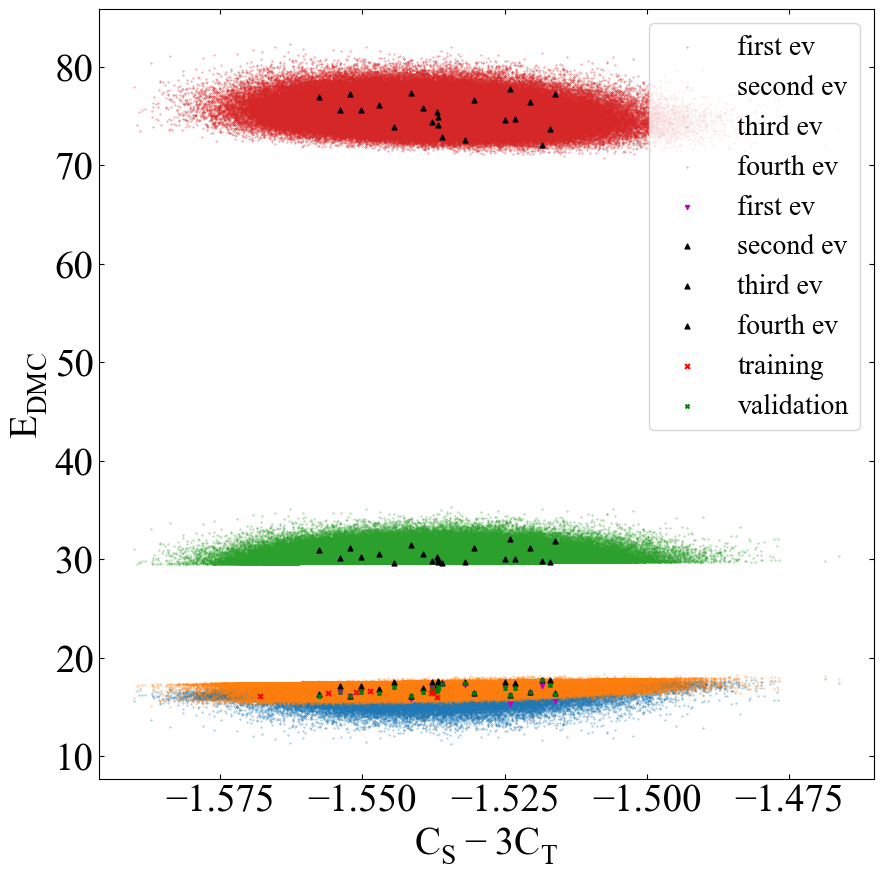}
\includegraphics[width=0.32\columnwidth]{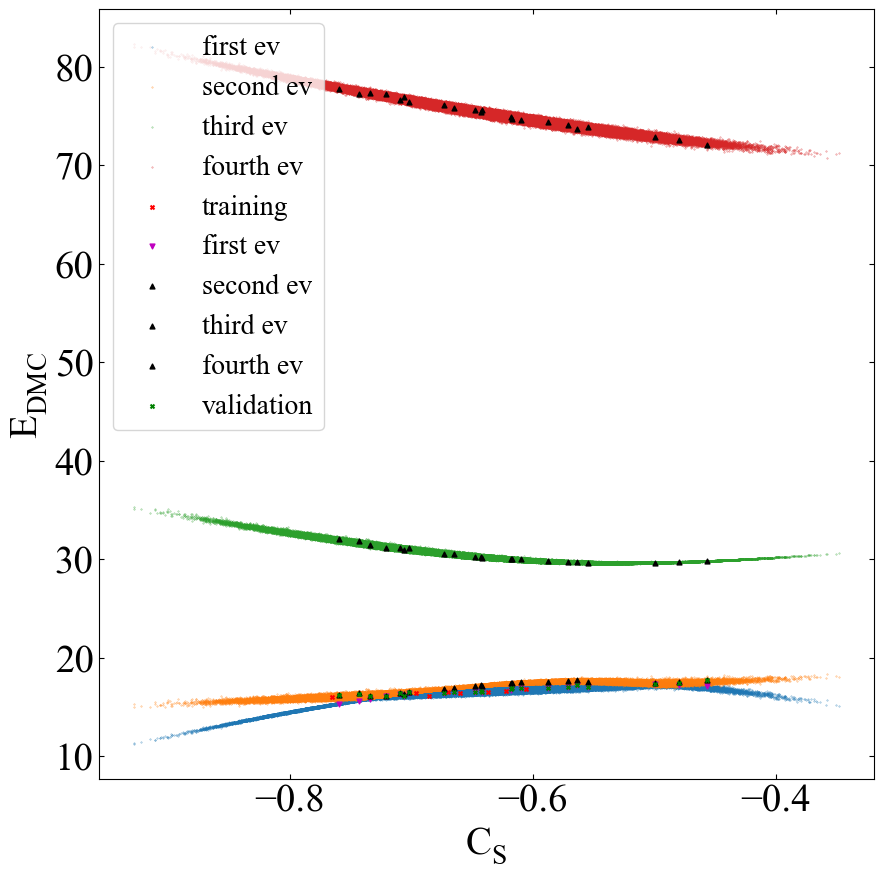}
\includegraphics[width=0.32\columnwidth]{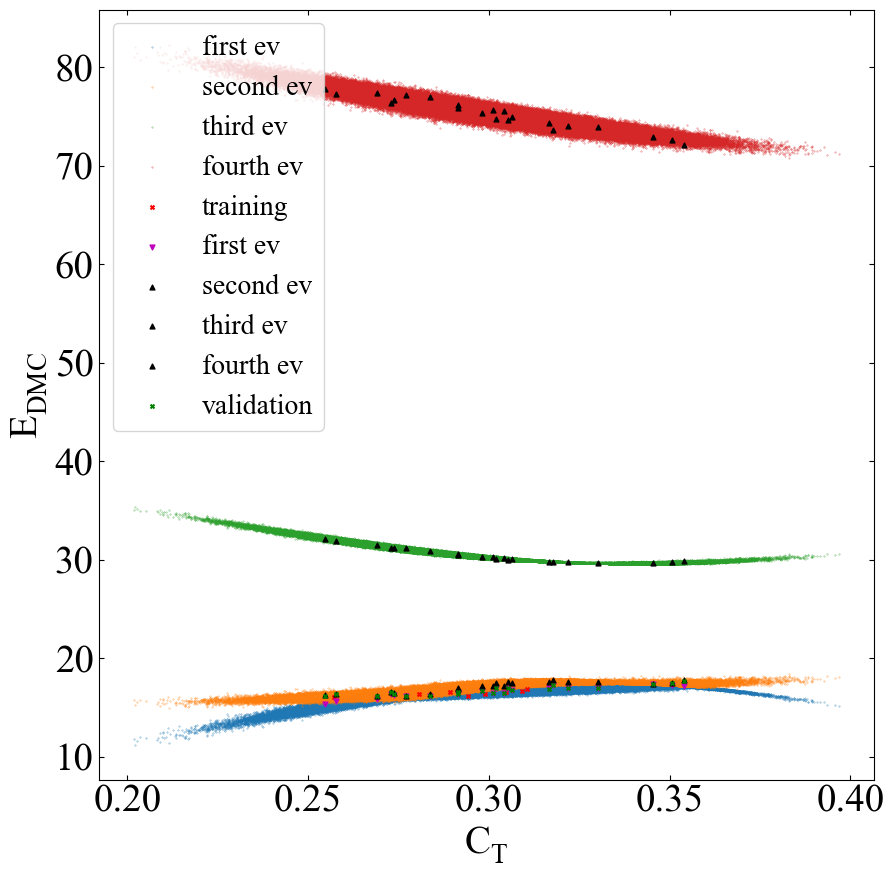}
    \caption{Same as Fig.~\ref{fig:upper_limits} but for N$_{\rm dim}=3$ (upper row) and N$_{\rm dim}=4$ (lower row). 
    }
    \label{fig:upper_limits_3_dim}
\end{figure*}

As discussed in the main text, at LO we generated 30 high-fidelity data points.
Initially, we randomly split them into a training set with 20 data points and a validation set with 10 validation points.
We then trained a two-dimensional PMM for various N$_{\rm train}$ keeping the random order of training points fixed and obtained the results in Fig.~\ref{fig:order_comparison}.
We find that the relative error decreases to sub-percent uncertainties for N$_{\rm train}=6$ but at larger values of training points the error can suddenly increase to much larger values, sometimes even to $20-30$\% at higher orders.
We found that this is due to level crossings of the two lowest eigenvalues of the PMM matrix that appear at energies above the ones included in the training set but intersects with the validation set, see Fig.~\ref{fig:upper_limits}.
In these cases, the correct PMM prediction would sometimes be given by the second eigenvalue, and the relative uncertainties for these validation points blow up the average uncertainty.
This can happen even if one additional training point is included and the PMM fit shifts accordingly, leading the the spikes observed in Fig.~\ref{fig:order_comparison}.
This problem also appears at larger N$_{\rm dim}$, see Figure~\ref{fig:upper_limits_3_dim} for N$_{\rm dim}=3$ and N$_{\rm dim}=4$. 

We then explored several options to avoid these level crossings.
A reasonable solution is to choose training points so that they span the full energy range. 
The simplest option is to include the highest-energy AFDMC result in the training set at first position, which effectively pushes higher eigenvalues to high energies for all N$_{\rm train}$. 
From Fig.~\ref{fig:order_comparison}, we see this this eliminates the issue and relative errors level off at very low values already for N$_{\rm train}\geq 4$.
When generating training data for the PMM, we cannot {\it a priori} ensure that we sample parameter sets that lead to sufficiently high energies.
Hence, in a second approach we first randomly split the data set into a training 
and a validation set, as before, but move the highest-energy datum in the training set on position one of the training set, i.e., it is always included in training. 
This leads to the green curve, partially reducing the spikes in relative error. 
Finally, as these spikes are related to the fit result for the matrix elements and suddenly appear at certain number of training data, it is reasonable to try to avoid them by starting the fit for N$_{\rm train}+1$ from the previous fit result for N$_{\rm train}$ for which the level crossing did not appear.
Testing this (red line), we find a similarly stable behavior as for the fit that always includes the global energy maximum, albeit with slightly larger uncertainties. 
Finally, we combined this approach with the one always including the training set maximum and obtain the purple curve. 
This approach leads to stable results with small errors, and we have decided to employ this to generate the PMMs in this paper.
            
\begin{figure*}[t]
    \centering
    \includegraphics[width=0.49\columnwidth]{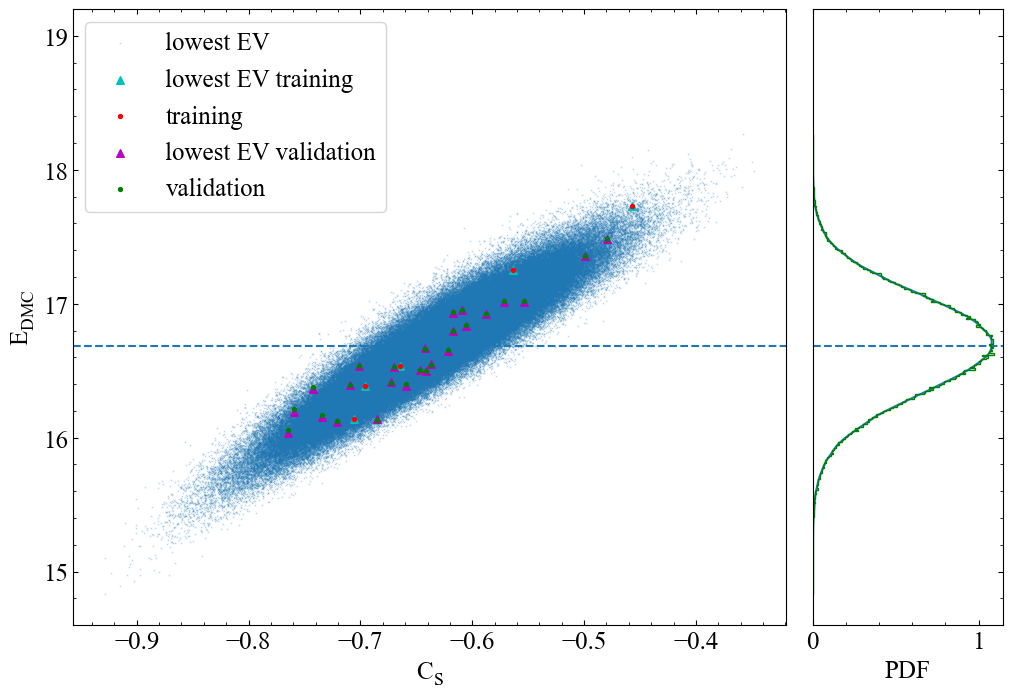} \hfil
    \includegraphics[width=0.49\columnwidth]{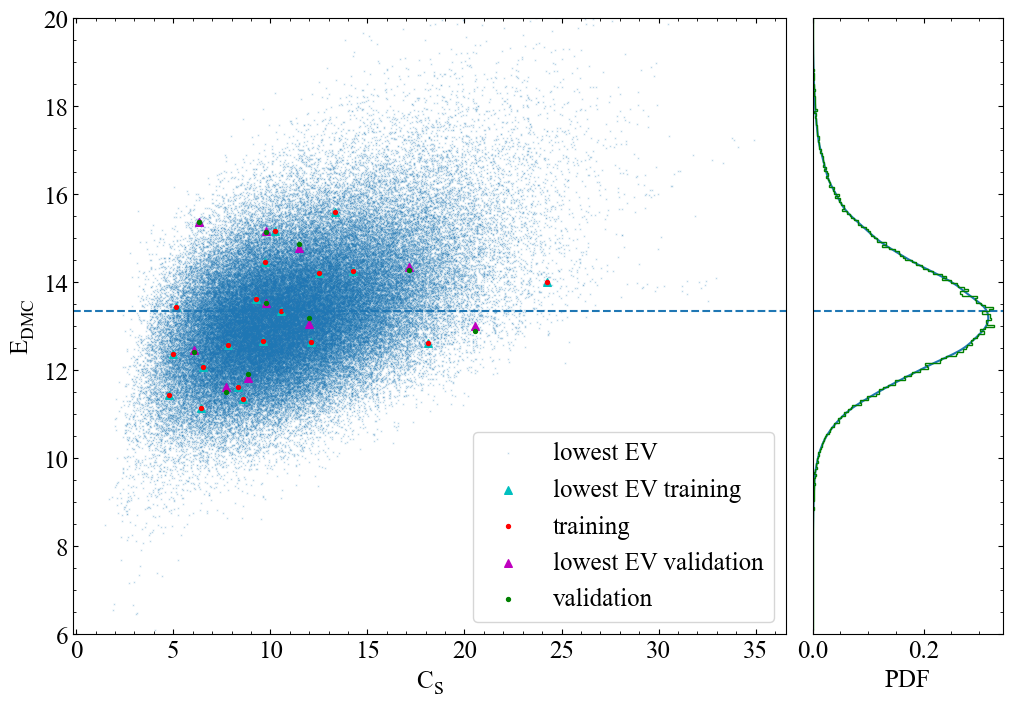}
    \caption{Left: Distribution of energies per particle at LO using a PMM fit with N$_{\rm train}=5$ and N$_{\rm dim}=2$.
    The distribution is obtained by propagating uncertainties on $C_S$ and $C_T$ to the observable.
    We also show the 1-D histogram and its Gaussian KDE fit.
    The average error for this fit is 0.3\%.
    Right: Similar, but for NLO with N$_{\rm train}=20$ and N$_{\rm dim}=3$.
    The average error for this fit is 0.5\%.
    }
    \label{fig:LO-NLO_distribution}
\end{figure*}

We have then employed this approach at LO at nuclear saturation density.
Using a PMM with 5 training points and N$_{\rm dim}=2$, we have propagated $\sim 200,000$~LEC samples to the energy per particle, see Fig.~\ref{fig:LO-NLO_distribution}.
We found that the final distribution changes only negligibly for different N$_{\rm train}$ and larger N$_{\rm dim}$.
We find that the uncertainty distribution at LO is well described by a Gaussian.

\subsection{PMM at NLO}

Finally, we discuss our PMM at NLO. 
We have extensively tested the PMM at NLO as the number of LECs at this order is the same as at N$^2$LO. 
At NLO, we have followed the same fit prescription as at LO.
As discussed in the main text, we have set up the NLO PMM in terms of 6 spectral LECs~\cite{Somasundaram:2023sup} to reduce the number of fit parameters. 
We have explicitly checked that the final errors are similar to PMMs defined in terms of 9 operator LECs, but the PMM convergence is faster and more stable.
In Fig.~\ref{fig:LO-NLO_distribution}, we show the distribution of energies per particle when propagating $\sim 200,000$~LEC sets at NLO. 
We find that the uncertainty distribution at NLO is also well described by a Gaussian. 
We conjecture that the different shape at N$^2$LO is due to the impact of three-nucleon forces.

\begin{figure*}[h]
    \centering
    \includegraphics[width=0.49\columnwidth]{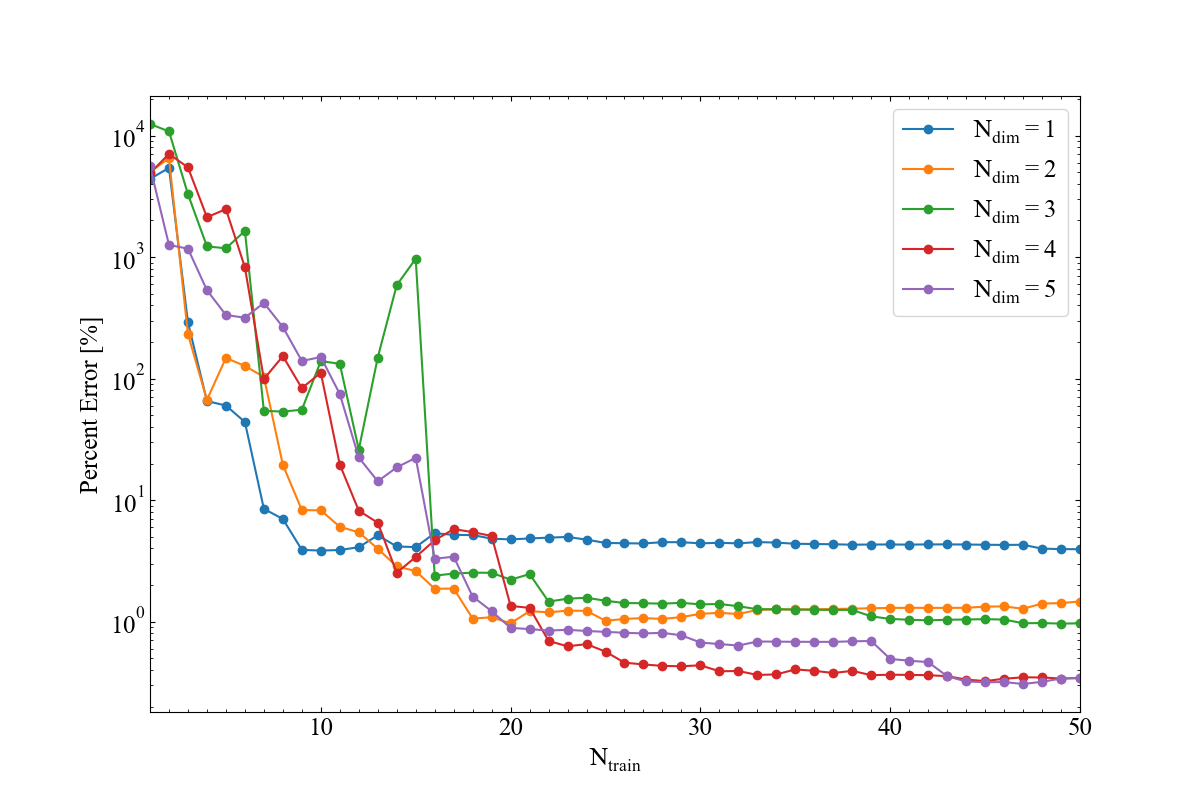}
    \includegraphics[
    width=0.49\columnwidth]{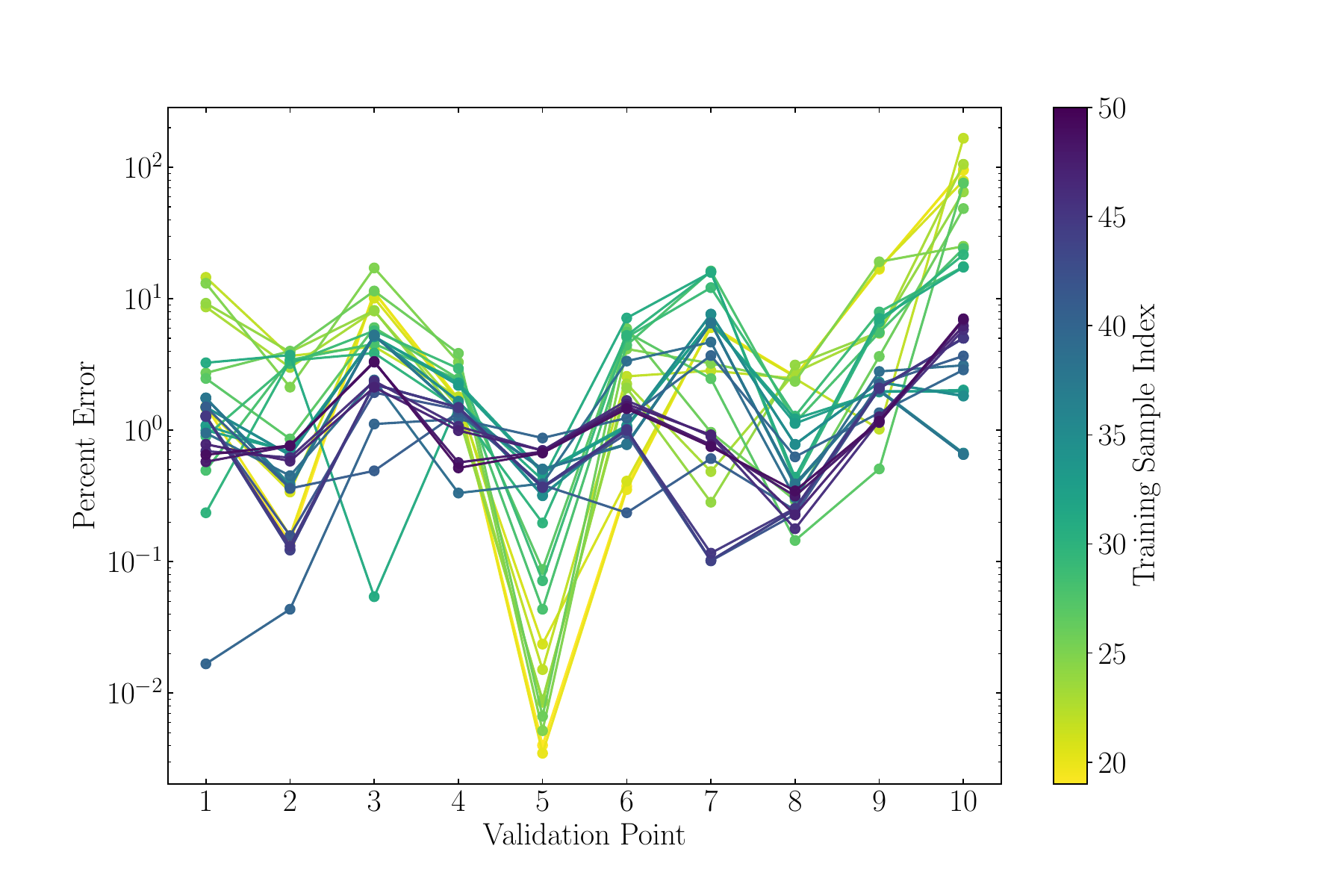}
    \caption{Left: The relative error of the PMM reproduction of the AFDMC validation energies at NLO as function of number of training points N$_{\rm train}$ for N$_{\rm dim}=1-5$ at nuclear saturation density. 
    The training and validation sets are the same for all N$_{\rm dim}$.
    Right: Percent errors for the 10 validation points for the PMM fits at NLO with N$_{\rm dim}=2$ and N$_{\rm train}=20-50$.
    }
    \label{fig:errors_validation_v2}
\end{figure*}

We show the convergence of the relative validation error as function of N$_{\rm train}$ in the left panel of Fig.~\ref{fig:errors_validation_v2}.
In contrast to the LO case, we find that the convergence is slower, and stabilizes for $\sim 20-25$ training points. 
We also find that the error improves up to N$_{\rm dim}=4$ but generally the error is comparable for N$_{\rm dim}=2-5$ for N$_{\rm train}=25$ and is of the order of 1\%.

In the right panel of Fig.~\ref{fig:errors_validation_v2}, we show the detailed evolution of the validation errors for the two-dimensional PMM for 20-50 training points.
We find that the relative validation errors can vary widely between validation data points, but that the uncertainty stabilizes at rather similar values if more and more training points are included. 

\begin{figure}[t]
    \centering
   \includegraphics[trim= 1.8cm 0.5cm 2.0cm 0.5cm, clip=, width=0.5\columnwidth]{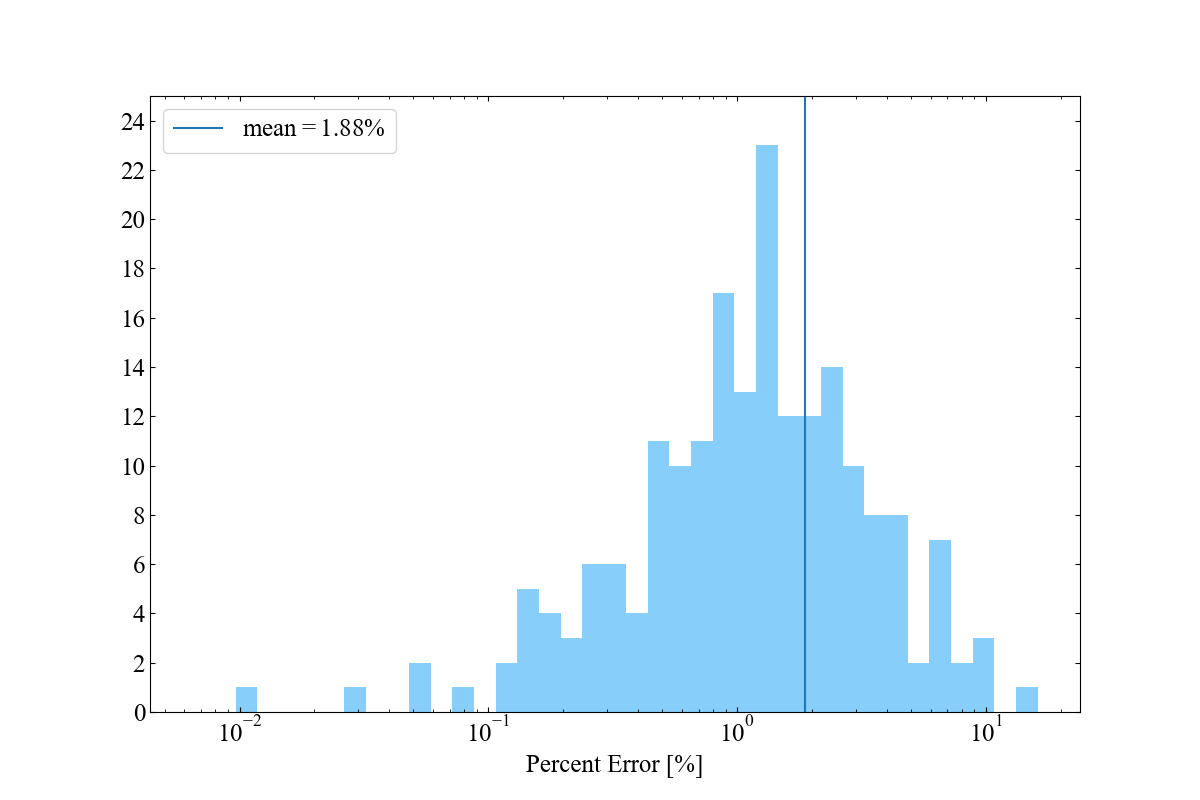}
    \caption{Distribution of PMM validation errors for all validation points at NLO using 6 spectral LECs and Dual Annealing and N$_{\rm dim} = 3$. 
    We have performed 20 different PMM fits for N$_{\rm train}=25$ randomly selected training points and 10 randomly selected validation points.
    }
    \label{fig:n_train_comparison}
\end{figure}

Finally, during the construction of our PMMs we have realized that for different random starting guesses of the matrix elements and for different sets of training and validation data, that the quality of the fit, i.e., the final relative validation error, can vary slightly. 
To quantify the variation, we have performed 20 different PMM fits for the NLO data. 
For each of these fits, we selected 10 random validation points and present a histogram of the validation error for all validation points in Fig.~\ref{fig:n_train_comparison}.
We find that the average error over all 20 PMMs is about $1.9$\%, which is close to the error in Fig.~\ref{fig:errors_validation_v2}.

\end{document}